\documentclass[sigplan,screen]{acmart}

\usepackage{xargs}
\usepackage{todonotes}

\newcommand{\bfemph}[1]{{\textbf{\textit{#1}}}}

\usepackage{hyperref}
\usepackage{graphicx}
\usepackage{subcaption}
\usepackage{textcomp}
\usepackage{amsmath} 
\usepackage{amssymb} 
\usepackage{mathtools}
\usepackage{booktabs}
\usepackage{makecell}
\usepackage[linesnumbered,ruled]{algorithm2e}
\usepackage{balance}
\usepackage{adjustbox}
\usepackage{array}

\usepackage{listings}
\renewcommand\footnotetextcopyrightpermission[1]{} 
\begin{document}

\title{A Fast Analytical Model of Fully Associative Caches}         
  

\author{Tobias Gysi}
\affiliation{
  \institution{ETH Zurich}
  \country{Switzerland}
}
\email{tobias.gysi@inf.ethz.ch}  

\author{Tobias Grosser}
\affiliation{
  \institution{ETH Zurich}
  \country{Switzerland}
}
\email{tobias.grosser@inf.ethz.ch}  

\author{Laurin Brandner}
\affiliation{
  \institution{ETH Zurich}
  \country{Switzerland}
}
\email{laurinb@student.ethz.ch}  

\author{Torsten Hoefler}
\affiliation{
  \institution{ETH Zurich}
  \country{Switzerland}
}
\email{htor@inf.ethz.ch}


\begin{abstract}
While the cost of computation is an easy to understand local property, 
the cost of data movement on cached architectures depends on 
global state, does not compose, and is hard to predict.
As a result, programmers often fail to consider the cost of data movement.
Existing cache models and simulators provide the missing information but are computationally expensive.
We present a \bfemph{lightweight cache model} for fully associative caches with least recently used (LRU) replacement policy that gives fast and accurate results.
%
We count the cache misses without explicit enumeration of all memory 
accesses by using \bfemph{symbolic counting} techniques twice:
1) to derive the stack distance for
each memory access and 2) to count the memory accesses with stack distance larger than the cache size. While this technique seems infeasible in theory, due to non-linearities after the first round of counting, we show that the counting problems are sufficiently linear in practice.
Our cache model often computes the results within seconds and contrary to simulation the execution time is mostly problem size independent.
Our evaluation measures modeling \bfemph{errors below~0.6\%} on real hardware.
By providing accurate data placement information we enable memory hierarchy aware software development.
\end{abstract}

\ccsdesc[300]{Software and its engineering~Software performance}
\ccsdesc[300]{Software and its engineering~Compilers}

\keywords{static analysis, cache model, performance tool}

\maketitle

\begin{figure}
  \centering 
  \includegraphics[width=0.85\columnwidth]{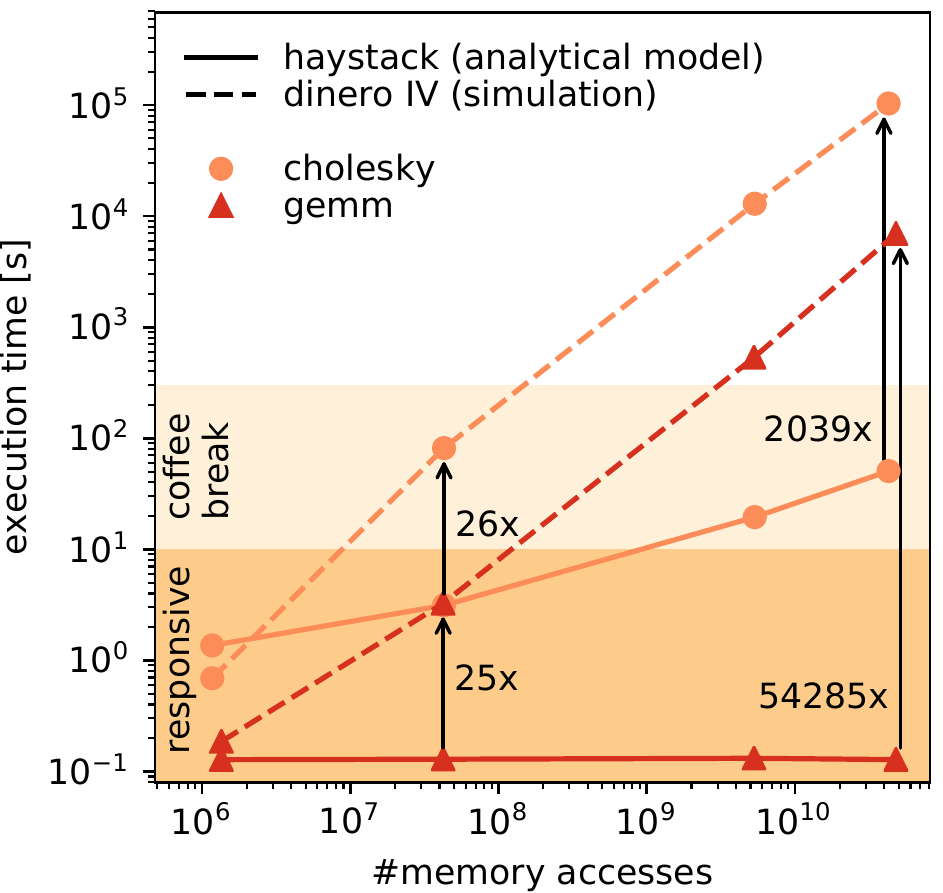}
  \caption{Scaling of the cache model compared to simulation.\label{fig-intro}}
\end{figure}

\section{Introduction}

Most programmers know the time complexity of their algorithms and tune codes by minimizing computation. Yet, ever increasing data-movement costs urge them to pay more attention to data-locality as a prerequisite for peak
performance.
When considering different implementation variants of an algorithm,
we typically have a good understanding of which variant performs
less computation or can be vectorized well. Selecting the optimal tile size or
deciding which loop fusion choice is optimal is far less intuitive. Essentially, 
we lack a perception of the cache state that allows us to reason about data movement.


Data-locality optimizations are often pushed to the end of the development
cycle when the code is available for benchmarking. But at this stage
eliminating fundamental design flaws may be hard. We believe a cache model
responsive enough to be part of the day-to-day workflow of a performance
engineer can provide the necessary guidance to make good design choices
upfront. After the completion of the development, the very same model
could provide the necessary data for accurate model driven automatic memory
tuning.

We present HayStack\footnote{\url{https://github.com/spcl/haystack}} the first cache model for fully associative caches with least recently used (LRU) replacement policy which is both fast and accurate. At the core of our model, we calculate the LRU stack distance~\cite{firststackdistance} (also called reuse distance~\cite{reusedistance, ding2003predicting, xiang2013hotl}) symbolically for each memory access.
The stack distance counts the distinct memory accesses between two subsequent accesses of the same memory location. All memory accesses with distance shorter than the cache size hit a fully associative LRU cache.

We show in \autoref{fig-intro} the scaling of HayStack compared to the Dinero~IV~\cite{dinero} cache simulator for increasing problem sizes. The simulation times are proportional to the problem size since simulators~\cite{dinero,casper,sniper,gem5} enumerate all memory accesses.
%
We use the Barvinok algorithm~\cite{barvinok} to count the cache misses. The algorithm avoids explicit enumeration by deriving symbolic expressions that evaluate to the cardinality of the counted affine integer sets and maps.
As demonstrated by the flat GEMM scaling curve, this symbolic counting makes the model execution time problem size independent. Even for Cholesky factorization, with its known non-linearities~\cite{epic} that prevent full symbolic counting, the scaling of the execution time remains flat compared to simulation.


While computing stack distances for static control programs is a well known
technique, reducing stack distance information for all dynamic memory accesses to a single cache miss count is difficult. Beyls et al.~\cite{epic} show that stack distances in general are non-affine. The divisions introduced when modeling cache lines add even more non-affine constraints. While symbolic summation over affine constraint sets is possible with the Barvinok algorithm, symbolic counting over non-affine constraints is considered hard in general. 

In this work, we show that this generally hard problem can in practice
become surprisingly tractable if non-linearities are carefully eliminated
by either specialization or partial enumeration. As a result we contribute:

\begin{itemize}
    \item The first efficient cache model to accurately predict static affine programs on fully associative LRU caches.
    \item An efficient hybrid algorithm that combines symbolic counting with partial enumeration to reduce the asymptotic cost of the cache miss counting.    
    \item A set of simplification techniques that exploit the regular patterns induced by the cache line structure to make the stack distance polynomials affine.
    \item An exhaustive evaluation which shows that our cache model performs well in practice with large speedups compared to existing approaches while achieving high accuracy compared to measurements on real hardware.
\end{itemize}

\section{Background}

We first introduce our hardware model, provide background on cache misses,
explain the concept of affine integer sets and maps, and discuss the set of considered programs.

\subsection{Hardware Model}

A cache implements various complex and sometimes undisclosed policies that 
define the exact behavior. We deliberately model a 
generic cache with full associativity and LRU replacement policy. 
When writing, we assume the caches allocate a cache line and load the memory
reference if necessary (write-allocate) and then forward the write to all
higher-level caches (write-through). We parametrize our cache model with the
cache line size $L$ and the cache size $C$ in bytes. When modeling multiple
cache hierarchy levels, we assume inclusive caches and specify the cache size
for every hierarchy level. These design choices avoid an overly detailed 
model that is only correct in a very controlled environment 
with know data alignment and allocation. As shown by \autoref{sec-acc}, 
we still model enough detail to produce actionable and accurate 
results in practice.

\subsection{Cache Misses}
\label{sec-cachemisses}

We assume that the modeled programs run in isolation and that their execution starts with an empty cache. We count data accesses and ignore instruction fetches.

According to Hill~\cite{cachemisstypes}, we distinguish three types of cache misses; 1) \emph{compulsory misses} happen if a program accesses a cache line for the first time, 2) \emph{capacity misses} happen if a program accesses too many distinct cache lines before accessing a cache line again, and 3) \emph{conflict misses} happen if a program accesses to many distinct cache lines that map to the same cache set of an associative cache before accessing a cache line again. We model fully associative caches and thus compute only compulsory and capacity misses. 


Not every access of a program variable translates in a cache access as the compiler may place scalar variables in registers. Compiler and hardware techniques such as out-of-order execution also change the order of the memory accesses. We assume all scalar variables are buffered in registers and count only array accesses in the order provided by the compiler front end. 

The cache misses measured when profiling a program depend on many factors generally unknown to an analytical cache model, for example, concurrent programs or the operating system may pollute the caches or the hardware prefetchers may load more data than necessary. We do not consider this system noise and instead provide an approximate but deterministic cache model.

\subsection{Integer Sets and Maps}

We use sets and maps of integer tuples to count the cache misses. We next define the relevant set and map operations necessary for the model implementation. These operations are a subset of the functionality provided by the integer set library (isl)~\cite{isl}. 

An affine set 
$$\textbf{S} = \{ (i_0,\dots,i_n) : con(i_0,\dots,i_n) \} $$ 
defines the subset of integer tuples $(i_0,\dots,i_n) \in \mathbb{Z}^n$ that satisfy the constraints $con(i_0,\dots,i_n)$. The constraints are Presburger formulas that combine affine expressions with comparison operators, boolean operators, and existential quantifiers. Presburger
arithmetic~\cite{complexity} also admits floor division and modulo with a constant divisor.

An affine map 
$$\textbf{R} = \{ (i_0,\dots,i_n) \rightarrow (j_0,\dots,j_m) : con(i_0,\dots,i_n, j_0,\dots,j_m) \} $$
defines the relation from integer tuples $(i_0,\dots,i_n) \in \mathbb{Z}^n$ to integer tuples $(j_0,\dots,j_m) \in \mathbb{Z}^m$ that satisfy the constraints $con(i_0,\dots,i_n, j_0,\dots,j_m)$ where the constraints have the same restrictions as the set constraints. The domain $\textbf{R}_{dom}$ defines the set of the integer tuples $(i_0,\dots,i_n)$ of the input dimensions for which a relation exists, and conversely the range $\textbf{R}_{ran}$ defines the set of integer tuples $(j_0,\dots,j_m)$ of the output dimensions for which a relation exists.

Both sets and maps support the set operations intersection $\textbf{S}_1 \cap \textbf{S}_2$, union $\textbf{S}_1 \cup \textbf{S}_2$, projection, and cardinality $|\textbf{S}|$. The domain intersection $\textbf{R} \cap_{dom} \textbf{S}$ intersects the domain of the map $\textbf{R}$ with the set $\textbf{S}$.
Maps also support the map operations composition $\textbf{R}_2 \circ \textbf{R}_1$ and inversion $\textbf{R}^{-1}$. 
The operator
\begin{align*}
\operatorname{lexmin}(\textbf{R}) = \{ (i_0, & \dots,i_n) \rightarrow (m_0,\dots,m_m) : \\
& \nexists (i_0,\dots,i_n) \rightarrow (j_0,\dots,j_m) \in \textbf{R}, \\
& \text{s.t.  } (j_0,\dots,j_m) \prec (m_0,\dots,m_m) \} 
\end{align*}
computes for every input tuple $(i_0,\dots,i_n)$ the lexicographic smallest output tuple $(m_0,\dots,m_m)$ of all tuples $(j_0,\dots,j_m)$ related to the input tuple.

A named set or map prefixes the integer tuples with names that convey semantic information. For example, we prefix the array element $\texttt{M}(2)$ with the array name and the statement instance $\texttt{S0}(1)$ with statement name. We use statement names starting with the letter $\texttt{S}$ and array names starting with any other letter. The names are semantically equivalent to an additional tuple dimension.

\subsection{Static Control Programs}
\label{sec-example}

\begin{figure}
\begin{lstlisting}
      int sum = 0;
      for(int i=0; i<4; ++i)
S0:    M[i] = i; 
      for(int j=0; j<4; ++j)
S1:    sum += M[3-j]; 
\end{lstlisting}
\caption{Example program used for illustration.}
\label{fig-example}
\end{figure} 

\begin{figure}
  \centering
  \includegraphics[width=\columnwidth]{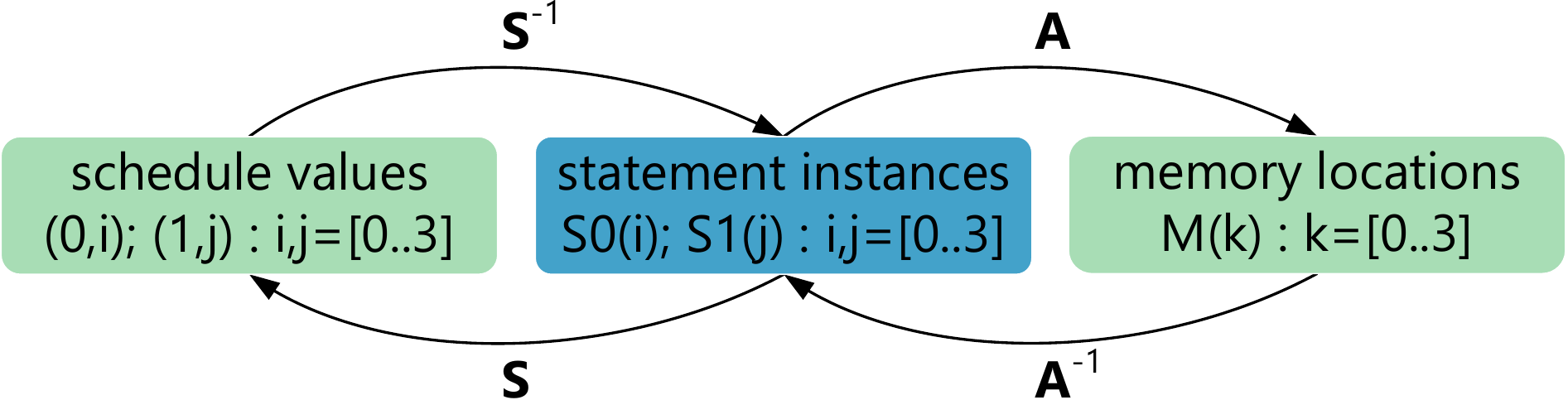}
    \caption{The statement instances and the related schedule values (schedule \textbf{S}) and memory accesses (access map \textbf{A}) are sufficient to compute the cache misses of a program.\label{fig-domains}}
\end{figure}

Our cache model analyzes affine \emph{static control programs} consisting of loop nests with known loop bounds that perform array accesses with affine index expressions. \autoref{fig-example} shows an example program with two statements: the statement~$\texttt{S0}$ initializes an array $\texttt{M}$ and the statement~$\texttt{S1}$ accumulates the array elements.
Before analyzing a program, we extract the sets and maps that specify the statement execution order and the memory access offsets. 

The iteration domain
$$ \textbf{I}=\{  \texttt{S0}(i) : 0 \leq i < 4; \texttt{S1}(j) : 0 \leq j < 4\} $$
defines the set of all executed statement instances. For the two statements of the example program, the loop variables $i$ and $j$ are limited to the range zero to three.
To define the execution order, the schedule
$$ \textbf{S} =\{ \texttt{S0}(i) \rightarrow (0,i) ;  \texttt{S1}(j) \rightarrow (1,j)\} \cap_{dom} \textbf{I} $$
maps the statement instances to a multi-dimensional schedule value. The statement instances then execute according to the lexicographic order of the schedule values. The intersection with the iteration domain $\textbf{I}$ limits the schedule domain to the program loop bounds. The access map
$$ \textbf{A}=\{ \texttt{S0}(i) \rightarrow \texttt{M}(i) ;  \texttt{S1}(j) \rightarrow \texttt{M}(3-j)\} $$
maps the array accesses of the statement instances to the accessed array elements.
The iteration domain $\textbf{I}$, the schedule $\textbf{S}$, and the access map $\textbf{A}$ capture all relevant program properties necessary to evaluate the cache model. 
\autoref{fig-domains} shows how the schedule $\textbf{S}$ and the access map $\textbf{A}$ relate statement instances, schedule values, and memory locations.

\section{Cache Model}

\begin{figure}
	\centering
	\includegraphics[width=\columnwidth]{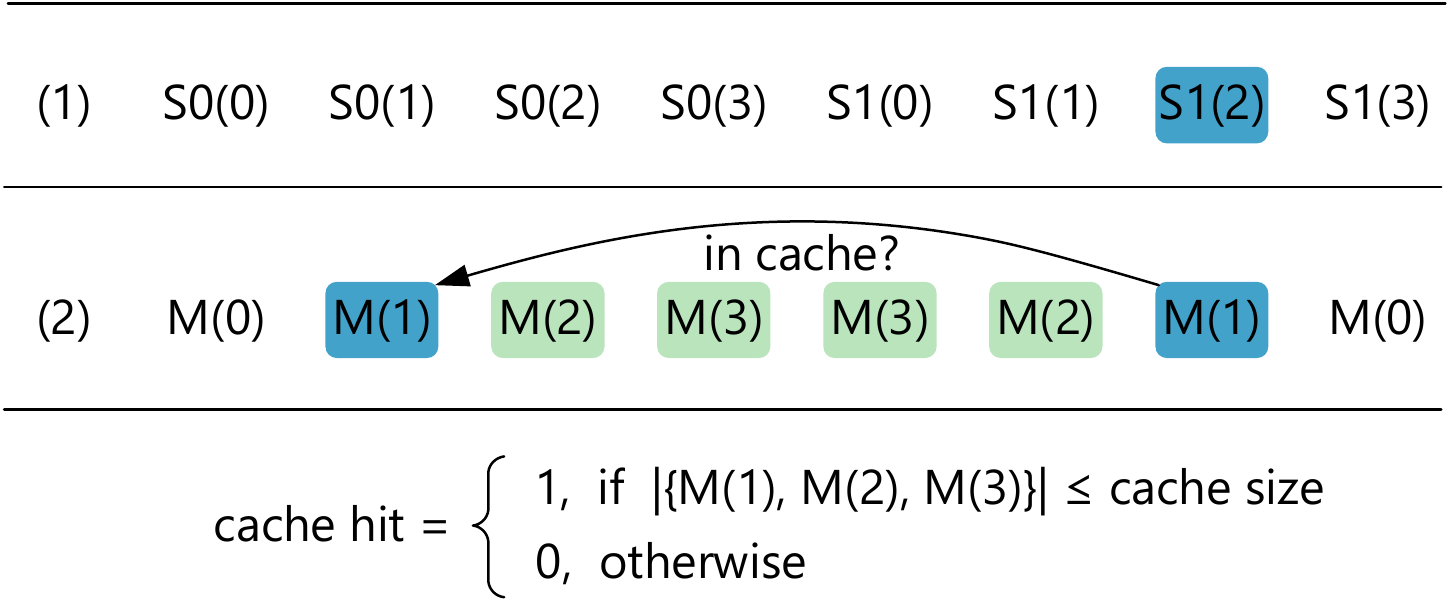}
    \caption{The (1) statement instance and the (2) memory access trace of the example program allow us to compute if the access $\texttt{M}(1)$ of the statement~$\texttt{S1}(2)$ hits the cache. \label{fig-stackdistance}}
\end{figure} 

Our cache model computes for every memory access the stack distance parametric in the loop variables and counts the instances with a stack distance larger than the cache capacity to determine the capacity misses.
All memory accesses with undefined backward stack distance access the cache line for the first time and count as compulsory misses.

\autoref{fig-stackdistance} shows the computation of the capacity misses for the example program introduced by \autoref{fig-example}: (1) enumerates the statement instances according to the schedule $\textbf{S}$ and (2) applies the access
map $\textbf{A}$ to the statement instances to compute the memory trace. Assuming the array element size is equal to the cache line size, the stack distance corresponds to the cardinality of the set $ \{ \texttt{M}(1), \texttt{M}(2), \texttt{M}(3) \} $ which contains the array elements accessed between and including the two subsequent accesses of $\texttt{M}(1)$. The second access of $\texttt{M}(1)$ hits the cache if the cardinality of the set is lower than or equal to the cache capacity.

\subsection{Computing the Stack Distance}
\label{sec-stackdistance}

The stack distance computation counts the number of distinct memory accesses between subsequent accesses of the same memory location. We determine for every memory reference the last access to the same memory location and count the set of memory accesses since this last access to obtain the stack distance parametric in the loop variables.

For our example program, the stack distance of the memory access in statement~$\texttt{S1}$ is equal to the loop variable $j$ plus one. We can thus express the
stack distance of the memory access with the map
$$\textbf{D} = \{ \texttt{S1}(j) \rightarrow j+1 : 0 \leq j < 4 \}$$
limited to the statement iteration domain. As the statement~$\texttt{S0}$ accesses all array elements for the first time its backward stack distance is undefined and the accesses count as compulsory misses. 

Our discussion of the stack distance computation initially assumes that every statement performs at most one access of a one-dimensional array with an element size equal to the cache line size. At the end of this section, we show how to overcome these limitations.

The memory accesses execute according to the statement execution order defined by the schedule.
The map
\begin{align*}
\textbf{L}_{\prec} = \{ (i_0, & \dots,i_n) \rightarrow (j_0,\dots,j_n) : \\
& (i_0,\dots,i_n) \prec (j_0,\dots,j_n) \land \\
& (i_0,\dots,i_n), (j_0,\dots,j_n) \in \textbf{S}_{ran} \} 
\end{align*}
relates the schedule values $(i_0,\dots,i_n)$ to all lexicographically larger schedule values $(j_0,\dots,j_n)$ and the map
\begin{align*}
  \textbf{L}_{\preceq} = \{ (i_0, & \dots,i_n) \rightarrow (j_0,\dots,j_n) : \\
  & (i_0,\dots,i_n) \preceq (j_0,\dots,j_n) \land \\
  & (i_0,\dots,i_n), (j_0,\dots,j_n) \in \textbf{S}_{ran} \} 
  \end{align*}
relates the schedule values $(i_0,\dots,i_n)$ to all lexicographically larger or equal schedule values $(j_0,\dots,j_n)$. Later on, we use these helper maps to filter relations by execution order.

The stack distance computation first identifies all accesses to the same array element. The equal map
$$ \textbf{E} = \textbf{S} \circ \textbf{A}^{-1} \circ \textbf{A} \circ \textbf{S}^{-1} $$
relates each schedule value to all schedule values that access the same array element. The concatenation $\textbf{A} \circ \textbf{S}^{-1}$ maps the schedule values to the accessed array elements and its reverse $\textbf{S} \circ \textbf{A}^{-1}$ maps the accesses back to the schedule values.
%
%
%
For our example program, the composition
\begin{align*}
\textbf{A} \circ \textbf{S}^{-1} = \{ & (0,i) \rightarrow \texttt{M}(i) : 0 \leq i < 4 ; \\
 & (1,j) \rightarrow \texttt{M}(3-j) : 0 \leq j < 4  \} 
\end{align*}
relates the schedule values to the accesses of the array $\texttt{M}$. The equal map then relates all schedule values that access the same array element. For example, the relations $(0,i)\rightarrow \texttt{M}(i)$ and $(1,j) \rightarrow \texttt{M}(3-j)$ access the same array element if $i$ is equal to $3-j$. The resulting equal map
\begin{align*}
\textbf{E} = \{ & (0,i) \rightarrow (0,i) :  0 \leq i < 4; \\
& (1,j) \rightarrow (1,j) :  0 \leq j < 4; \\
& (0,i) \rightarrow (1,j) : j=3-i \land 0 \leq i < 4; \\
& (1,j) \rightarrow (0,i) : i=3-j \land 0 \leq j < 4\}
\end{align*}
contains the relation $(0,i) \rightarrow (1,j)$ with $j=3-i$ and its reverse but also the self relations of the schedule values.

\begin{figure}
	\centering
	\includegraphics[width=\columnwidth]{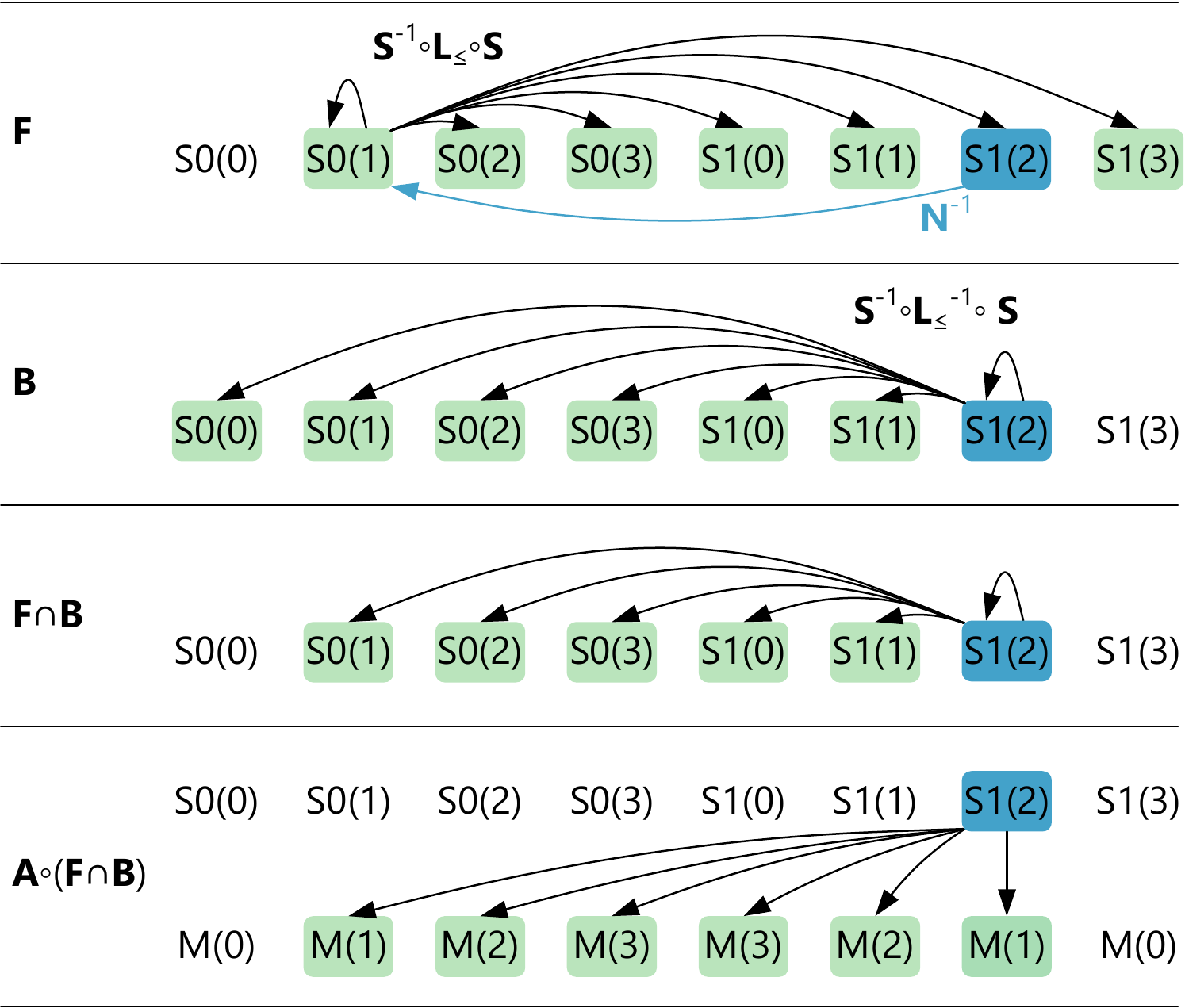}
    \caption{     
    The relations of the forward map $\textbf{F}$ and the backward map $\textbf{B}$ for the statement instance $\texttt{S1}(2)$ of the example program (the forward map $\textbf{F}$ corresponds to the concatenation of the blue backward arrow and the black forward arrows). The map intersection defines the statement instance between and including the two accesses of $\texttt{M}(1)$. The concatenation with the map $\textbf{A}$ yields the related memory accesses.    
    \label{fig-forwardbackward}}
\end{figure}  

The lexicographically shortest relations of the equal map denote the subsequent accesses to the same array element which are closest in time.
The next map
$$ \textbf{N} = \textbf{S}^{-1} \circ \operatorname{lexmin}(\textbf{L}_{\prec} \cap \textbf{E}) \circ \textbf{S} $$
intersects the equal map $\textbf{E}$ with the map $\textbf{L}_{\prec}$ to filter out all backward in time and self relations and the $\operatorname{lexmin}$ operator removes all forward in time relations except for the shortest ones. We compose the result with $\textbf{S}$ and $\textbf{S}^{-1}$ to convert the schedule values to statement instances. The next map consequently relates every statement instance to the next statement instance that accesses the same array element. 
For our example program, the equal map contains only the forward relation $(0,i) \rightarrow (1,j)$ which means the $\operatorname{lexmin}$ operator has no effect since there is only one relation per statement instance. The next map
$$\textbf{N} = \{ \texttt{S0}(i) \rightarrow \texttt{S1}(j) : j=3-i \land 0 \leq i < 4 \}$$
thus relates the instances of statement~$\texttt{S0}$ to the instances of statement~$\texttt{S1}$ that access the same array element.

The next map contains subsequent statement instances that access the 
same array element but not the statement instances executed in between. To compute them, we intersect the set of statement instances executed after the first access with the set of statement instances executed before the second access of the same array element. \autoref{fig-forwardbackward} illustrates this intersection.
The backward map
$$ \textbf{B} = \textbf{S}^{-1} \circ \textbf{L}_{\preceq}^{-1} \circ \textbf{S} $$
relates the statement instances to all statement instances with lexicographically smaller or equal schedule value. The maps $\textbf{S}$ and $\textbf{S}^{-1}$ convert from statement instances to schedule values and back. The forward map
$$ \textbf{F} = (\textbf{S}^{-1} \circ \textbf{L}_{\preceq} \circ \textbf{S}) \circ \textbf{N}^{-1}   $$
relates the statement instances to all statement instances with lexicographically larger or equal schedule value than the statement instance that last accessed the same array element. We reverse the next map $\textbf{N}$ to compute the statement instance that accessed the array element last. The intersection of the forward map and the backward map contains all statement instances executed between subsequent accesses of the same array element.

\autoref{fig-forwardbackward} shows the forward and backward map relations for the statement instance $\texttt{S1}(2)$ of the example program that accesses the array element $\texttt{M}(1)$. The forward map $\textbf{F}$ corresponds to the concatenation of the blue backward arrow and the black forward arrows. The intersection of the two maps contains the statement instances executed between the subsequent accesses of the array element $\texttt{M}(1)$. We finally concatenate this intersection with the access map $\textbf{A}$ to obtain the stack distance map that relates every statement instance to the array accesses performed since the last access of the same array element.

The number of related array elements defines the stack distance of the statement instances in the stack distance map.
We use the isl~\cite{isl} implementation of the Barvinok algorithm~\cite{barvinok} to count the relations symbolically. The algorithm computes the map cardinality by counting the points of the range related to every point of the domain. The result of the computation are quasi polynomials parametric in the input dimensions of the map that evaluate to the number of related range points. As the domain is not always homogeneous, the algorithm splits the map domain into pieces that consist of a quasi polynomial and the subdomain of the map domain where the polynomial is valid. After counting the stack distance map, the distance set
$$ \textbf{D} = \{| \textbf{A} \circ (\textbf{F} \cap \textbf{B}) |\} $$
contains pieces with quasi polynomials parametric in the schedule input dimensions that for a subdomain of the iteration domain evaluate to the stack distance. The pieces do not overlap and together cover the full iteration domain. For our example program, the distance set 
$$\textbf{D} = \{ \texttt{S1}(j) \rightarrow j+1 : 0 \leq j < 4 \} $$
contains one piece with the polynomial $\texttt{S1}(j) \rightarrow j + 1$ and the domain $0 \leq j < 4$ covering the entire iteration domain.

%
 
\paragraph{cache lines and multi-dimensional arrays}
An adapted access map $\textbf{A}$ that relates statement instances to cache lines instead of array elements suffices to support cache lines and multi-dimensional arrays. Let us assume our example program initializes the diagonal elements of a two-dimensional array $\texttt{M}(i,i)$. Then the access map 
$$ \textbf{A} = \{ \texttt{S0}(i) \rightarrow \texttt{M}(i,c = \lfloor i * E / L\rfloor) \} $$
models the accessed cache lines given the size of the array elements $E$ and cache line size $L$ in bytes. We replace the innermost dimension of the array access with the cache line index $c$, which multiplies the array index with the element size and divides the result by the cache line size. As a result, accesses of neighboring array elements map to the same cache line. The outer dimensions of the array index remain unchanged since we assume the innermost dimension is cache line aligned and padded to an integer multiple of the cache line size. This restriction can be lifted at the expense of a more complex formulation.

\paragraph{multiple memory accesses per statement}
An extension of the schedule $\textbf{S}$ and the access map $\textbf{A}$ with an additional schedule dimension that orders the memory accesses of the statements allows us to model more than one memory access per statement. Let us assume the statement~$\texttt{S0}$ of the example program reads the array element $\texttt{I}(i)$ and writes the result to the array element $\texttt{M}(i)$. We then extend the schedule
$$ \textbf{S} = \{ \texttt{S0}(i,a) \rightarrow (0,i,a); \texttt{S1}(j,a) \rightarrow (1,j,a) \} $$ 
with the access dimension $a$ that orders the memory accesses of the statement. Then the access map
$$ \textbf{A} = \{ \texttt{S0}(i,0) \rightarrow \texttt{I}(i);  \texttt{S0}(i,1) \rightarrow \texttt{M}(i); \texttt{S1}(j,0) \rightarrow \texttt{M}(3-j) \} $$  
assigns every array access to a unique statement instance since the access dimension enumerates the array accesses of every statement in the order provided by the compiler front end. The extended schedule executes only one array access per statement instance and thus requires no further modifications of the stack distance computation.



\medskip
The output of the stack distance computation is a set of polynomials that defines the backward stack distance for every array access of the static control program.

\subsection{Counting the Capacity Misses}
\label{sec-capacitymisses}

All memory accesses with stack distance larger then the cache size count as capacity miss. As discussed in \autoref{sec-stackdistance}, the stack distance computation splits the iteration domain into pieces. Each piece defines the stack distance for a subdomain of the iteration domain. To obtain the capacity misses, we count for every piece the points of the subdomain for which the polynomial evaluates to a stack distance larger than the cache size.

The piece with polynomial $\texttt{S1}(j) \rightarrow j + 1$ and domain $0 \leq j < 4$ defines the stack distance for the entire iteration domain of our example program. The cache miss set
$$\textbf{M} = \{ \texttt{S1}(j) :  j + 1 > C \land 0 \leq j < 4 \} $$
contains all points of the piece with stack distance larger than cache size $C$ which means the cardinality of the cache miss set $|\textbf{M}|$ is equal to the number of capacity misses. Assuming cache size two, the cache miss set contains the statement instances $\texttt{S1}(2)$ and $\texttt{S1}(3)$ that cause two capacity misses. 

The distance set specifies the stack distance for all program statements. To count the capacity misses per statement, we split the distance set by statement and compute the cache misses separately. Without loss of generality, we discuss the cache miss computation for a statement~$\texttt{S0}$.

\begin{figure}
	\centering
	\includegraphics[width=\columnwidth]{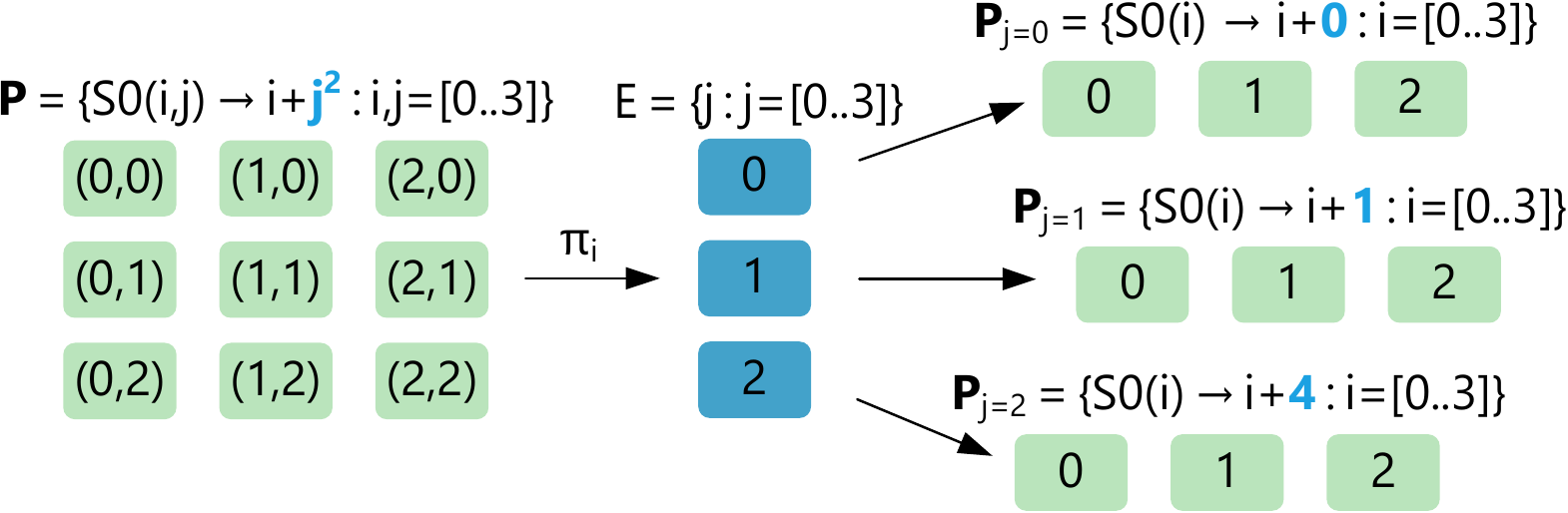}
    \caption{To count the non-affine piece $\textbf{P}$, we project out the affine $i$-dimension to obtain the enumeration domain $\textbf{E}$. We next bind the $j$-dimension of the piece $\textbf{P}$ to the $j$-values in the enumeration domain and separately count the cache misses for the resulting affine pieces $\textbf{P}_{j=0}$, $\textbf{P}_{j=1}$, and $\textbf{P}_{j=2}$.\label{fig-enumerate}}
\end{figure}  

The Barvinok algorithm also computes the set cardinality by counting the points symbolically.
We use the algorithm to count affine cache miss sets and resort to explicit enumeration for non-affine sets. As explicit enumeration is expensive, we only enumerate the non-affine polynomial dimensions and count the affine dimensions symbolically. This \emph{partial enumeration} technique splits cache miss sets into pieces with affine lower-dimensional polynomials. \autoref{fig-enumerate} demonstrates the technique for an example polynomial with non-affine $j$-dimension. \autoref{sec-tuning} discusses further techniques to split non-affine pieces into multiple affine pieces.

\begin{algorithm}
  \DontPrintSemicolon
  \SetKwData{CacheMisses}{T}
  \SetKwData{CacheSize}{C}
  \SetKwData{Distances}{\textbf{D}}
  \SetKwData{Piece}{\textbf{P}}
  \SetKwData{Enumerate}{\textbf{E}}
  \SetKwData{Point}{pt}
  \SetKwData{FixedPiece}{\textbf{P}\textsubscript{pt}}
  
  \SetKwFunction{IsPieceAffine}{isPieceAffine}
  \SetKwFunction{CountAffinePiece}{countAffinePiece}
  \SetKwFunction{GetEnumerationDomain}{getNonAffineDomain}
  \SetKwFunction{fixEnumerationDimensions}{bindNonAffineDimensions}

  \SetKwInOut{Input}{input}
  \SetKwInOut{Output}{output}
  \SetKwInOut{Parameter}{parameter}
  
  \Input{\Distances distance set of pieces}
  \Output{\CacheMisses total number of cache misses}
  \Parameter{\CacheSize cache size}
  \BlankLine
  \CacheMisses $\leftarrow$ 0\;
  \ForEach{\Piece in \Distances}{ \label{alg-forbeg}
    \eIf{\IsPieceAffine{\Piece} } {  \label{alg-affinebeg}
      \CacheMisses$\leftarrow \CacheMisses + \CountAffinePiece{\Piece, \CacheSize}$\;  \label{alg-affineend}
    }{
      \Enumerate$\leftarrow$\GetEnumerationDomain{\Piece}\; \label{alg-nonaffinebeg}
      \ForEach{\Point in \Enumerate}{ 
        \FixedPiece$\leftarrow$\fixEnumerationDimensions{\Piece, \Point}\; \label{alg-fix}
        \CacheMisses$\leftarrow \CacheMisses + \CountAffinePiece{\FixedPiece, \CacheSize}$\; \label{alg-countfix}  \label{alg-nonaffineend}     
      }
    }
  } \label{alg-forend}
  \Return \CacheMisses\;
  \caption{counting the capacity misses}\label{alg-counting}
\end{algorithm}

\autoref{alg-counting} counts the total number of cache misses T given the distance set $\textbf{D}$ of the program. The algorithm enumerates all pieces $\textbf{P}$ of the distance set (lines \ref{alg-forbeg}-\ref{alg-forend}). Every piece $\textbf{P}$ consists of a polynomial and a domain that define the stack distance of a memory access for a subdomain of the iteration domain. If the polynomial of the piece $\textbf{P}$ is affine we count the cache misses symbolically (lines \ref{alg-affinebeg}-\ref{alg-affineend}), otherwise the \emph{partial enumeration} projects the non-affine dimensions out of the domain of the piece $\textbf{P}$ and enumerates all points of the resulting non-affine enumeration domain $\textbf{E}$ (lines \ref{alg-nonaffinebeg}-\ref{alg-nonaffineend}). For every such point pt, we bind the non-affine dimensions of the piece $\textbf{P}$ to the coordinates of the point pt and count the cache misses of the affine piece \textbf{P}\textsubscript{pt} symbolically. \autoref{fig-enumerate} illustrates the splitting of non-affine pieces (lines \ref{alg-nonaffinebeg}-\ref{alg-nonaffineend}).  

The method \emph{countAffinePiece} counts the cache misses of the piece $\textbf{P}$ with affine stack distance polynomial. 
A polynomial is affine if its degree is zero or one. We first compute the cache miss set
$$ \textbf{M} = \{ \texttt{S0}(i_0,\dots,i_n) : \textbf{P}_p(i_0,\dots,i_n) > C \land (i_0,\dots,i_n) \in \textbf{P}_D \} $$
where $\textbf{P}_p$ denotes the polynomial and  $\textbf{P}_D$ the domain of the piece $\textbf{P}$. The cache miss set contains all memory accesses with stack distance larger than cache size $C$. To count the cache misses, we compute the cardinality $|\textbf{M}|$ using the Barvinok algorithm.

The method \emph{getNonAffineDomain} projects all points of the piece $\textbf{P}$ to the non-affine dimensions to obtain the enumeration domain $\textbf{E}$.
For example, \autoref{fig-enumerate} projects the piece 
$$\textbf{P} = \{ \texttt{S0}(i,j) \rightarrow i + j^2 : 0 \leq i < 3 \land 0 \leq j < 3 \} $$
which contains the quadratic term $j^2$. We project the points to the non-affine $j$-dimension to compute the enumeration domain $\textbf{E} = \{ j : 0 \leq j < 3 \} $. The enumeration always spans all dimensions with degree larger than one. But the polynomial may also contain product terms with multiple dimensions. We then greedily select the dimensions that conflict with most other dimensions. For example, if the polynomial contains the products $ij$ and $ik$ we enumerate the $i$-dimension since it conflicts with both other dimensions.

The method \emph{bindNonAffineDimensions} binds the non-affine dimensions of the piece $\textbf{P}$ to the values of the point pt.
For example, \autoref{fig-enumerate} binds the $j$-dimension of the piece 
$$\textbf{P} = \{ \texttt{S0}(i,j) \rightarrow i + j^2 : 0 \leq i < 3 \land 0 \leq j < 3 \} $$
to the value two and obtains the piece
$$\textbf{P}_{j=2} = \{ \texttt{S0}(i) \rightarrow i + 4 : 0 \leq i < 3 \} $$
which we can count with the method \emph{countAffinePiece}.

The counting algorithm works for all static control programs and avoids complete enumeration except all dimensions are non-affine.
 
\subsection{Eliminating Non-Affine Terms}
\label{sec-tuning}

Many stack distance polynomials contain non-affine terms that prevent fast symbolic counting. We develop rewrite strategies that eliminate non-affine terms containing floor expressions. The floor expressions themselves are quasi-affine but often appear in products with other non-constant operands modeling effects such as the stack distance variation for different cache line offsets. We specialize the stack distance polynomials for different cache line offsets to make them affine which enables the efficient symbolic counting.



\begin{figure}
	\centering
  \includegraphics[width=0.95\columnwidth]{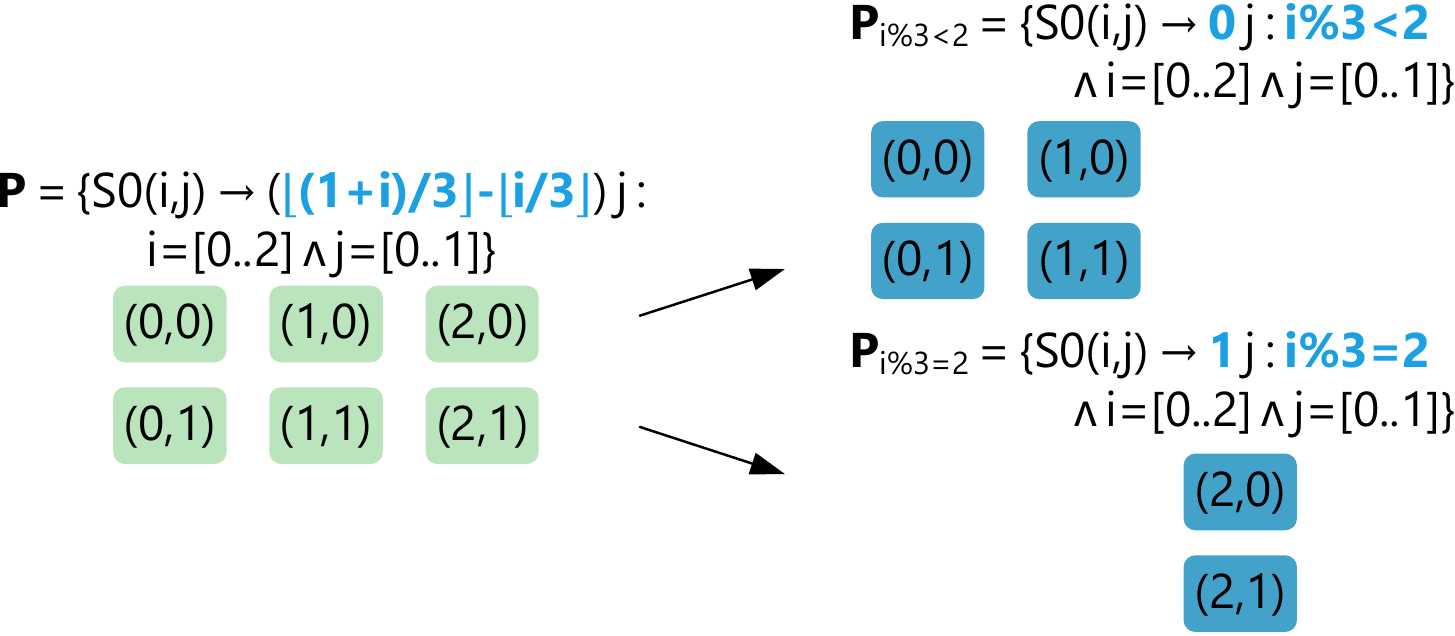}
    \caption{\emph{Equalization} replaces the non-affine piece~$\textbf{P}$ with the affine pieces $\textbf{P}_{i\%3<2}$ and $\textbf{P}_{i\%3=2}$ to model a stack distance that varies at the last cache line offset.
    \label{fig-equalization}} 
  \bigbreak
	\centering
	\includegraphics[width=0.95\columnwidth]{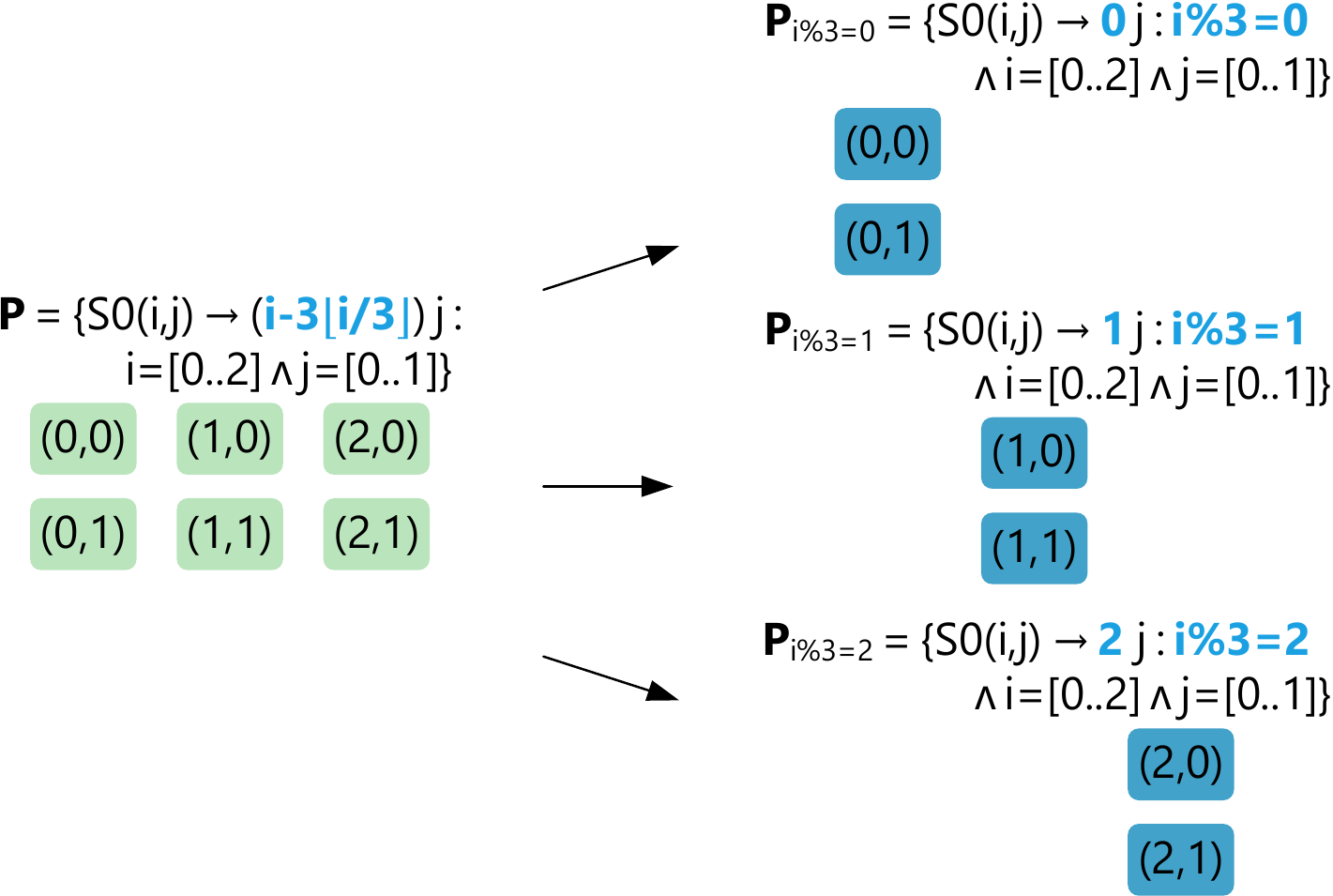}
    \caption{\emph{Rasterization} replaces the non-affine piece~$\textbf{P}$ with the affine pieces $\textbf{P}_{i\%3=0}$, $\textbf{P}_{i\%3=1}$, and $\textbf{P}_{i\%3=2}$ to model a stack distance that varies at every cache line offset.\label{fig-rasterization}} 
\end{figure}  



The floor expressions of some polynomials differ only by a constant offset. For example, the piece 
$$\textbf{P}=\{ \texttt{S0}(i,j) \leftarrow (\left\lfloor (1+i)/3 \right\rfloor - \left\lfloor i /3 \right\rfloor)j : 0 \leq i < 3 \land 0 \leq j < 2 \} $$
contains the floor expressions $\left\lfloor (1+i)/3 \right\rfloor$ and $\left\lfloor (i)/3 \right\rfloor$.
The two floor expressions are equal except if $i$ modulo three is equal to two. Then the second floor expression is larger by one.
The difference of the two floor expressions thus evaluates to zero for the first two elements and to one for the last element of every cache line. \autoref{fig-equalization} shows how to introduce simplified polynomials for the first two and the last element of every cache line. This \emph{equalization} technique splits the cache line in multiple regions that typically contain more than one element. 

The polynomials may also contain terms with the plain variable and other terms which compute the floor of the variable. For example, the piece 
$$ \textbf{P}=\{ S0(i,j) \rightarrow ( i - 3 \left\lfloor i/3 \right\rfloor )j : 0 \leq i < 3 \land 0 \leq j < 2 \} $$
contains the floor expression $3 \left\lfloor i/3 \right\rfloor$ which is equal to $i$ except for a constant that depends on the cache line offset. \autoref{fig-rasterization} shows how to replace the polynomial with one simplified polynomial per cache line offset. This \emph{rasterization} technique enumerates all cache line offsets.

We apply the two floor elimination techniques in the order of presentation and only keep the results if the degree of at least one simplified polynomial is lower than the degree of the original polynomial.

\begin{figure*}
  \centering
    \begin{subfigure}{\columnwidth}
    \centering
    \includegraphics[width=\columnwidth]{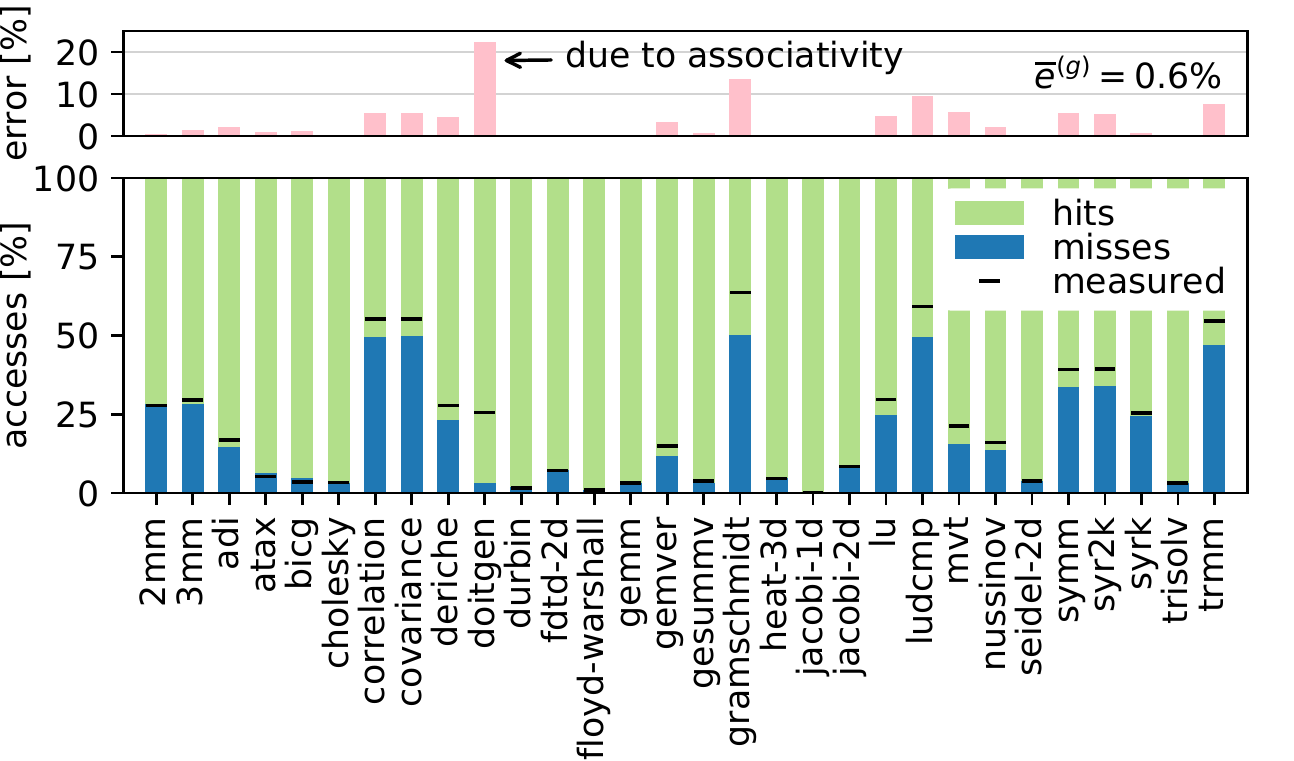}
    \caption{L1 cache}
    \end{subfigure}\hfill
  \centering
    \begin{subfigure}{\columnwidth}
    \centering
    \includegraphics[width=\columnwidth]{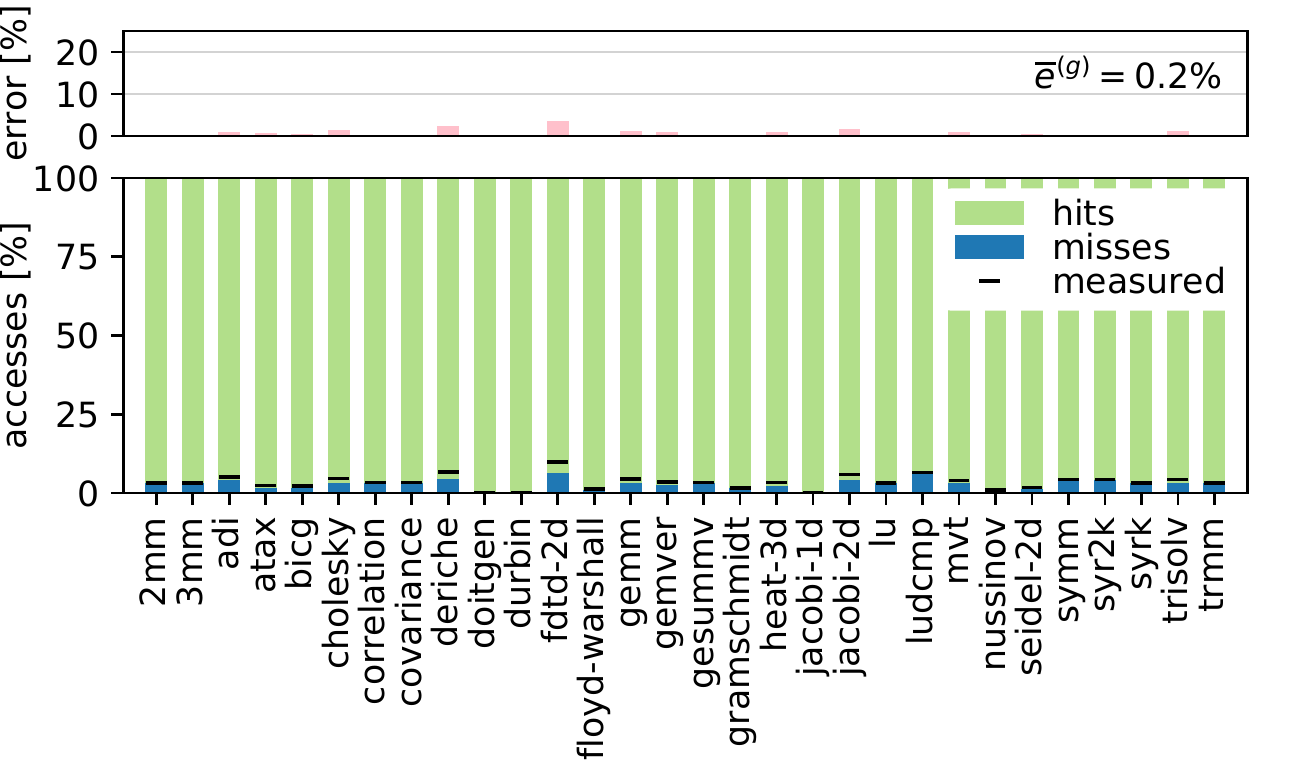}
    \caption{L2 cache}
    \end{subfigure}\hfill
    \caption{Cache misses and hits predicted by HayStack compared to the measured cache misses (median of 10 measurements) for the PolyBench kernels with the prediction error relative to the number of memory accesses on top.} 
    \label{fig-accmodel}
\end{figure*}
\begin{figure*}
    \begin{subfigure}{\columnwidth}
    \centering
    \includegraphics[width=\columnwidth]{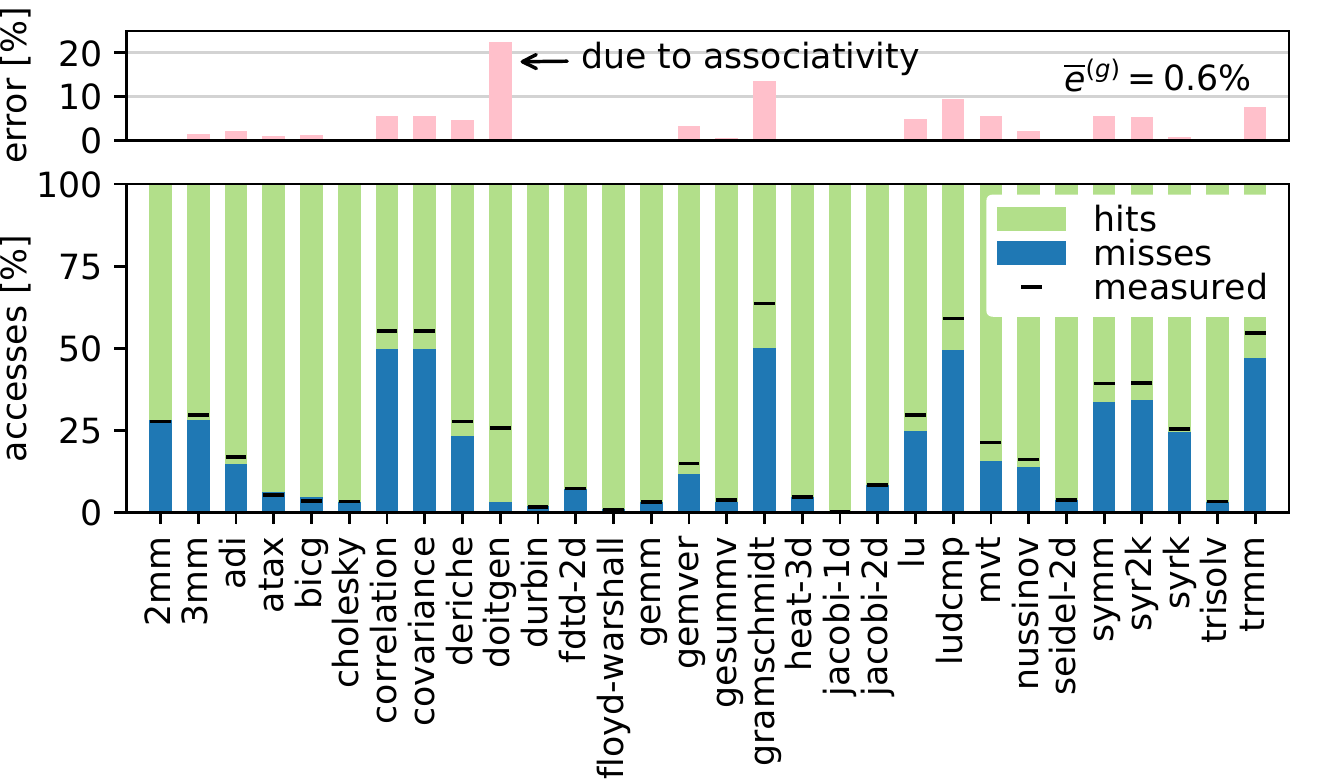}
    \caption{L1 cache (fully associative)}
    \end{subfigure}\hfill
    \begin{subfigure}{\columnwidth}
    \centering
    \includegraphics[width=\columnwidth]{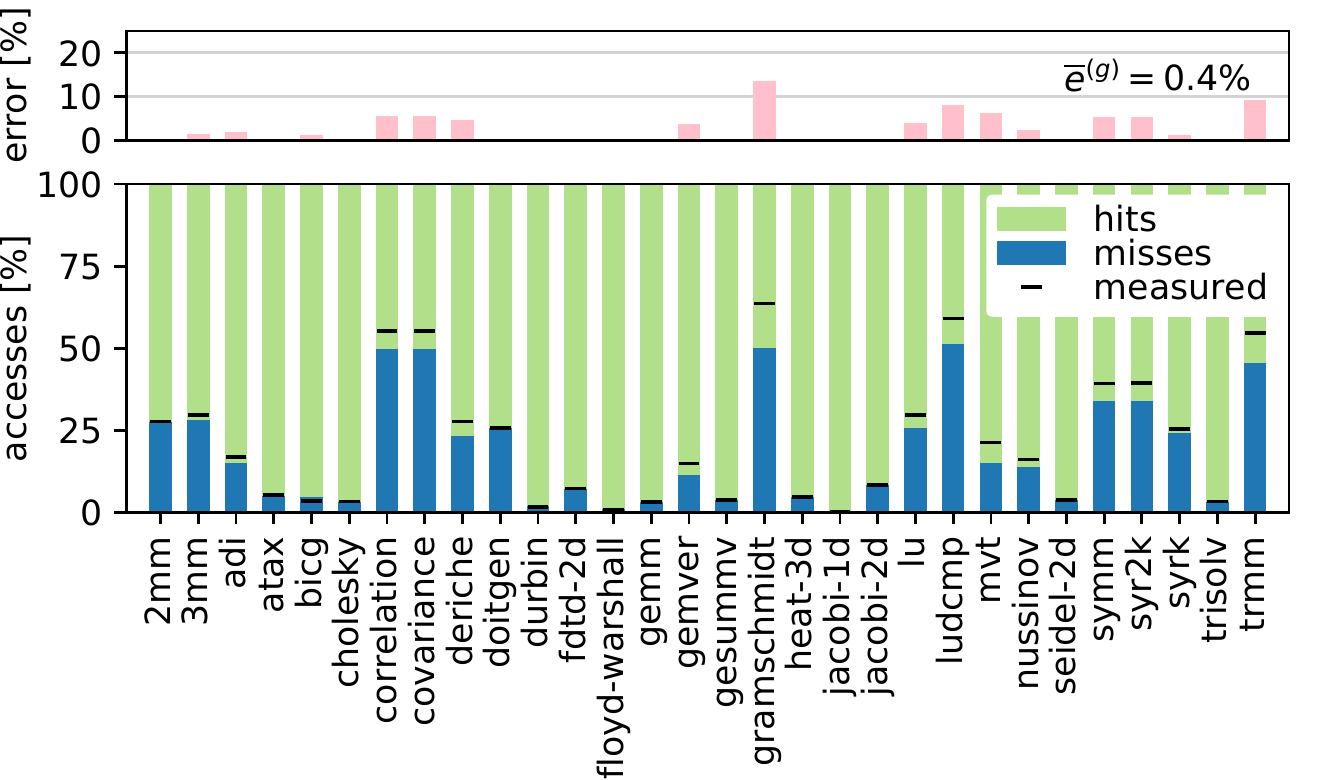}
    \caption{L1 cache (8-way associative)}
    \end{subfigure}\hfill
    \caption{Cache misses and hits simulated by Dinero~IV compared to the measured cache misses (median of 10 measurements) for the PolyBench kernels with the prediction error relative to the number of memory accesses on top.} 
    \label{fig-accdinero}
\end{figure*}

\subsection{Counting the Compulsory Misses}

All memory accesses that touch a cache line for the first time are compulsory misses. 

As the array $\texttt{M}$ of our example program is initialized by the statement~$\texttt{S0}$, the first map
$$\textbf{F} = \{ \texttt{M}(i) \rightarrow \texttt{S0}(i) : 0 \leq i < 4 \} $$
relates every array element to the statement instance that accesses the element first which means the cardinality $|\textbf{F}_{dom}|$ of the first map domain counts the compulsory misses.

The compulsory misses are the memory accesses with lexicographically minimal schedule value. The first map
$$\textbf{F} = \textbf{S}^{-1} \circ\operatorname{lexmin}(\textbf{S} \circ \textbf{A}^{-1})$$
thus selects for every memory access the lexicographically minimal relation of the composition $\textbf{S} \circ \textbf{A}^{-1}$ that relates memory accesses to schedule values and composes the result with the inverse schedule $\textbf{S}^{-1}$ to obtain the related statement instances. The composition with the inverse schedule allows us to intersect the range of the first map with the iteration domain of the individual statements to count the compulsory misses per statement.
For our example program, the composition
\begin{align*}
\textbf{S} \circ \textbf{A}^{-1} = \{ & \texttt{M}(i) \rightarrow (0,i) : 0 \leq i < 4; \\
& \texttt{M}(j) \rightarrow (1,3-j) : 0 \leq j < 4\}
\end{align*}
contains two accesses for every array element. The $\operatorname{lexmin}$ operator removes the second access due to the lexicographically larger schedule value. After the composition with the inverse schedule $\textbf{S}^{-1}$, we use the Barvinok algorithm to count the compulsory misses $|\textbf{F}_{dom}|$. 




\subsection{Computational Complexity}

All compute-heavy parts of our cache model perform Presburger arithmetic that in general is known to have very high computational complexity~\cite{complexity, nguyen}. 
The established complexity bounds range from polynomial time decidable~\cite{lenstra} for expressions with fixed dimensionality and only existential quantification to double exponential~\cite{fischer} for arbitrary expressions. Haase~\cite{complexity} presents further results that show a complexity increase with the dimensionality and the number of quantifier alternations of the Presburger expression.

The Presburger relations computed by our cache model have only existential quantification and the dimensionality is limited by the loop depth suggesting polynomial complexity. 
Yet, the cache model may introduce further variables to model divisions or modulo operations making the complexity exponential in the number of dimensions.

Although the cache model has exponential worst-case complexity, the empirical performance evaluation presented in \autoref{sec-perf} shows that our cache model performs well for typical input programs. The dimensionality of the observed Presburger relations remains limited since most real-world programs do not make extensive use of branch conditions and index expressions that result in integer divisions or modulo operations.








\section{Evaluation}
\label{sec-eval}

We next evaluate the performance of HayStack and compare its accuracy to simulated and measured results.

\subsection{Setup and Methodology}

We evaluate on a test system with two 18-core Intel Xeon Gold 6150 processors. Every core has a 32KiB L1 cache (8-way set associative) and an inclusive 1MiB L2 cache (16-way set associative). The non-inclusive 18x1.375MiB L3 cache (11-way set associative) is shared among all cores. A non-inclusive cache may and an inclusive cache has to duplicate all cache lines stored by the lower-level caches. All caches load the cache line before writing (write-allocate) and forward the write only if the cache line is evicted (write-back). 

We compile with GCC 6.3 and use the Dinero~IV cache simulator~\cite{dinero} to compute and the PAPI-C library~\cite{papi} to measure the number of cache misses.
We evaluate the model for a number of different kernels. PolyBench
4.2.1-beta~\cite{polybench} is a collection of static control programs that
implement algorithmic motifs from scientific computing. If not stated
otherwise the PolyBench experiments use the default configuration (large) and the model emulates fully associative L1 and L2 caches with the capacities of the test system. 

All performance measurements run single-threaded using only one core of the test system. To quantify measurement noise, the execution times show the median and the non-parametric 95\% confidence intervals~\cite{scientificbenchmarking} of 10 measurements.

\subsection{Accuracy Overview}
\label{sec-acc}

All mathematical models are a trade-off between accuracy and complexity. A static cache model cannot predict dynamic measurement noise for example due to concurrent code execution. We aim at an accurate prediction of the cache misses without modeling too many implementation details. 
 
A comparison to measurements on a real system is the main benchmark for every cache model. To measure the cache misses, we compile the PolyBench~\cite{polybench} kernels with PAPI~\cite{papi} support using GCC optimization level O2. PolyBench~\cite{polybench} flushes the caches before every kernel execution which allows us to measure compulsory and capacity misses. We collect the counters \texttt{PAPI\_L1\_DCM} and \texttt{PAPI\_L2\_DCM} that sum the data cache misses for the L1 and L2 caches, respectively. \autoref{fig-accmodel} compares the sum of the compulsory and capacity misses predicted by HayStack to the measured cache misses shown by black lines.
Most kernels cause more cache misses than predicted which is expected since we model idealized fully associative caches with LRU instead of pseudo-LRU replacement policy. We also do not consider possible overfetch due to the hardware prefetchers.
To quantify the error,
\autoref{fig-accmodel} shows for every kernel the prediction error relative to the total number of memory accesses computed by the model. Most kernels have low single digit prediction errors with a geometric mean error of 0.6\% and 0.2\% for the L1 cache and the L2 cache, respectively. Only doitgen and gramschmidt have prediction errors above 10\%. 

We also execute the PolyBench kernels with Dinero~IV~\cite{dinero} to simulate the number of cache misses with full associativity and with the associativity of our test system. \autoref{fig-accdinero} compares the sum of the simulated compulsory, capacity, and conflict misses to the measured cache misses shown by black lines. 
We observe that the simulation results for the fully associative L1 cache qualitatively agree with the model.
All simulation results are within 0.1\% of the model for the L1 cache and within 3\% of the model for the L2 cache (relative to the total number of memory accesses). We conclude that our design decisions of padding the innermost dimension of multi-dimensional arrays, discussed in \autoref{sec-stackdistance}, and modeling only array accesses and not scalar accesses, discussed in \autoref{sec-cachemisses}, have no significant impact on the accuracy of the model.
The simulation results with test system associativity eliminate the error for the doitgen kernel. We conclude that modeling set associativity is only relevant for one of the PolyBench kernels.
The error of the remaining kernels is dominated by other error sources such as the difference between LRU and pseudo-LRU replacement policy that are neither considered by the simulator nor by the model.

HayStack reproduces the simulation results for full associativity and the associativity mismatch compared to the test system does not dominate the modeling error. 

\subsection{Performance Overview}
\label{sec-perf}

We next analyze the performance of HayStack and its sensitivity to model parameters such as the problem size or the number of cache hierarchy levels.

\begin{figure}
	\centering
	\includegraphics[width=\columnwidth]{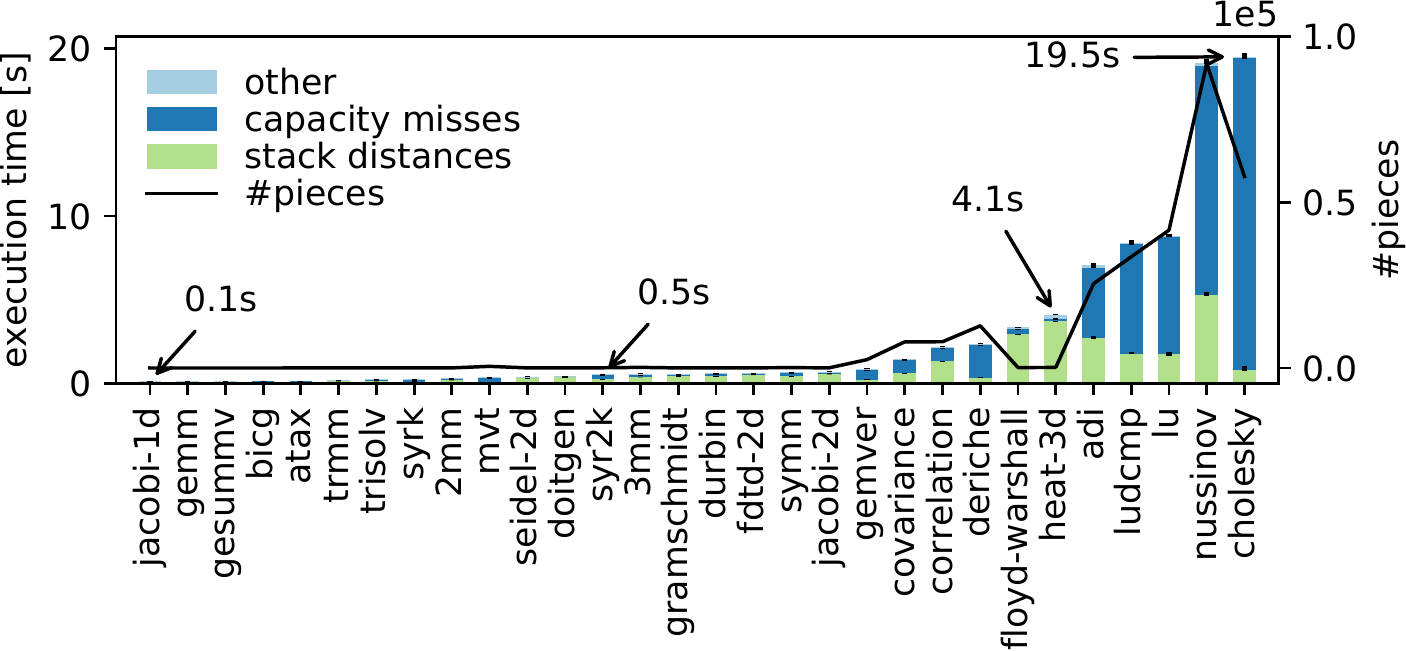}
    \caption{Execution times for the main components of HayStack compared to the number of separately counted pieces for the PolyBench kernels sorted by execution time. 
    \label{fig-benchcost}}
\end{figure}

\begin{figure}
	\centering
	\includegraphics[width=\columnwidth]{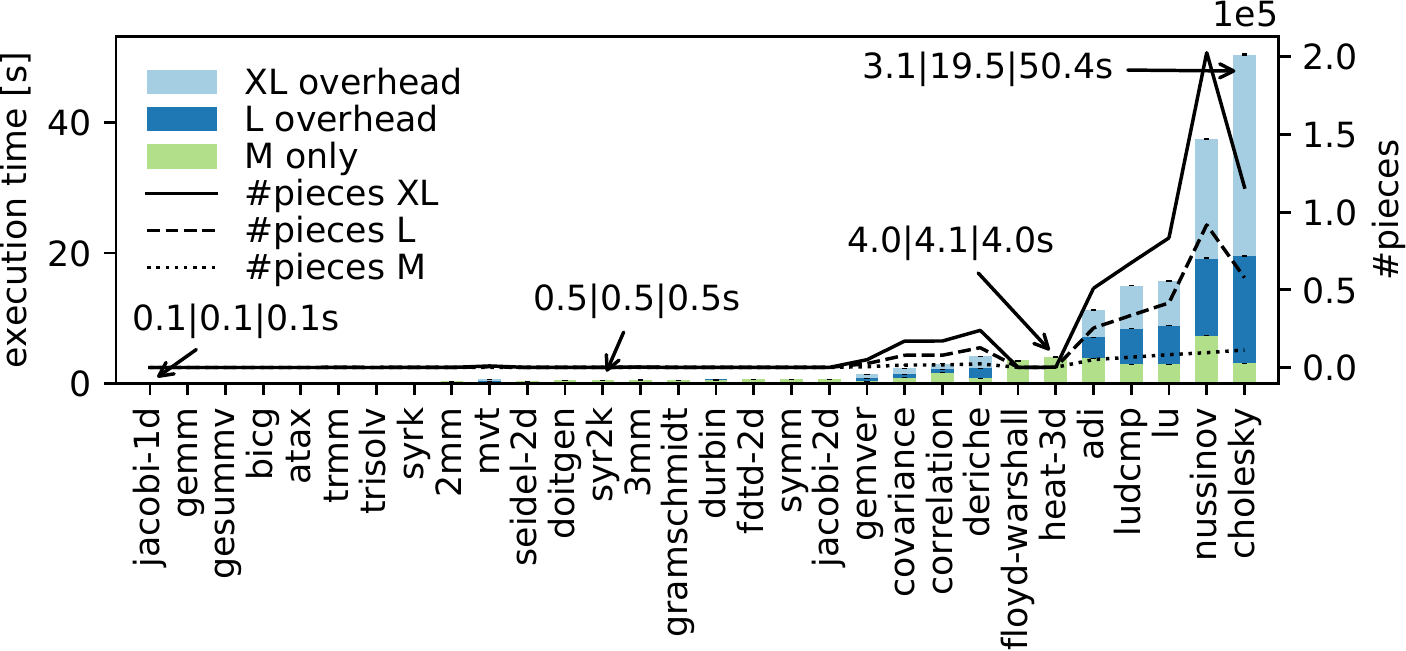}
    \caption{Execution times for the extra large (XL), large (L), and medium (M) problem sizes of PolyBench compared to the number of counted pieces.\label{fig-benchsizes}}
\end{figure}

Two components dominate the model execution time: 1) the stack distance computation discussed in \autoref{sec-stackdistance} and 2) the capacity miss counting discussed in \autoref{sec-capacitymisses}. \autoref{fig-benchcost} shows the cost of the two components compared to the total model execution times for the PolyBench kernels. The analysis of most kernels terminates within 5 seconds (jacobi-1d to heat-3d) while the more expensive kernels take up to 20 seconds (adi to cholesky). The capacity miss counting dominates the cost of the expensive kernels. When counting the capacity misses, the \emph{partial enumeration} and to a lesser extend the \emph{equalization} and \emph{rasterization}, discussed in \autoref{sec-tuning}, split the iteration domain into pieces with affine stack distance polynomials that support symbolic counting. The solid line in \autoref{fig-benchcost} shows the number of counted pieces. We observe that the expensive kernels require more splits due to non-affine stack distance polynomials and that the counting costs correlate with the number of pieces.

\begin{figure}
	\centering
  \includegraphics[width=\columnwidth]{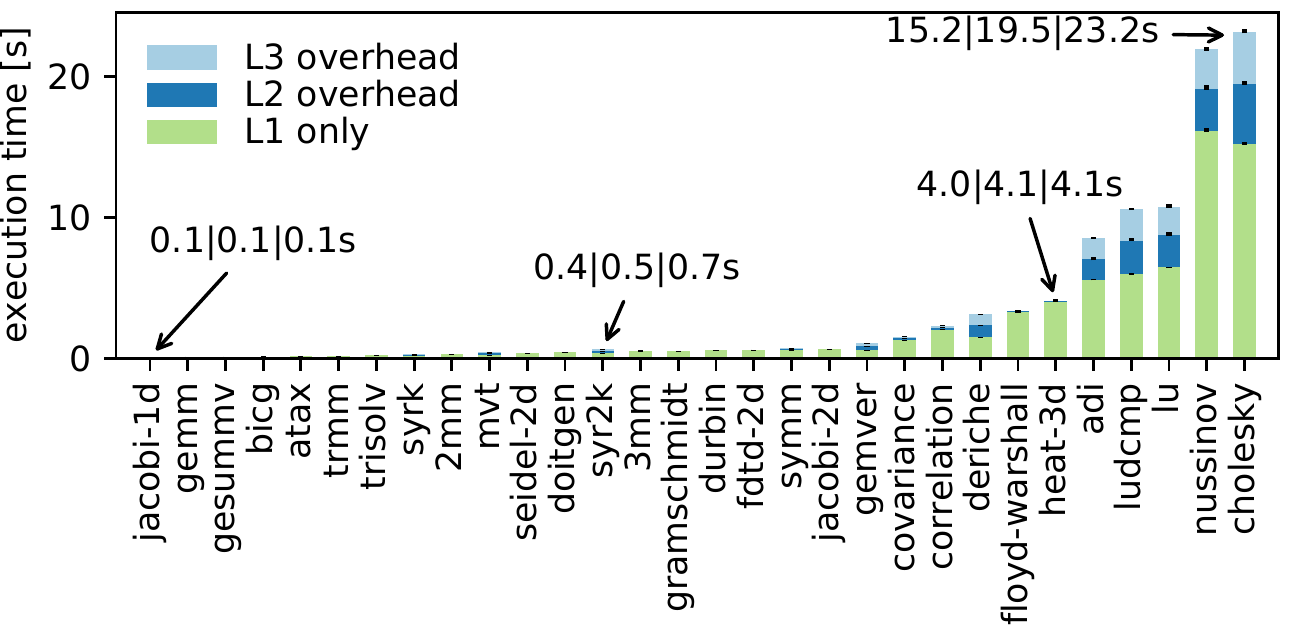}
  \caption{Comparison of the execution times when modeling one, two, or three cache hierarchy levels.\label{fig-benchcaches}}
\end{figure} 

\begin{figure} 
	\centering
	\includegraphics[width=\columnwidth]{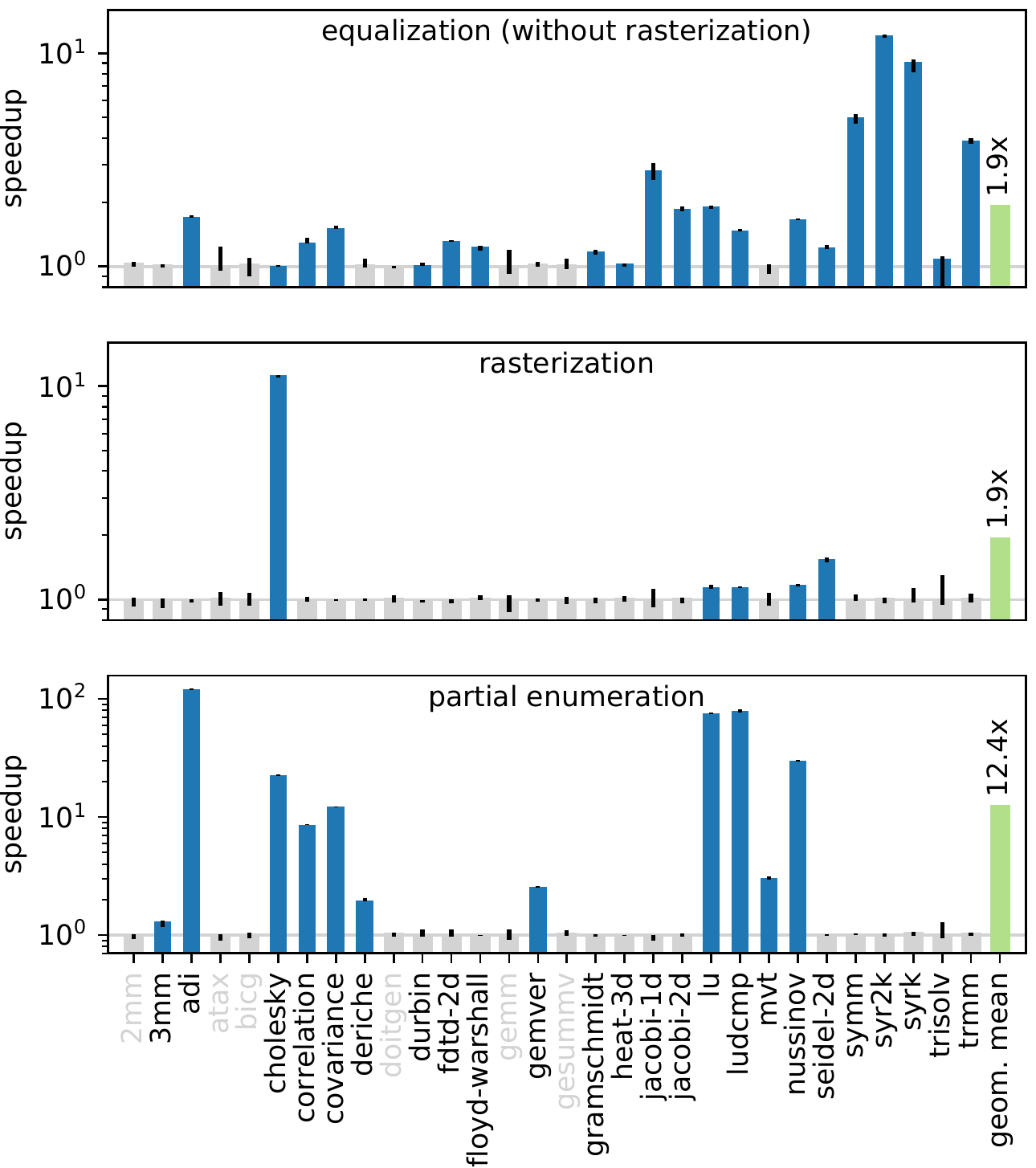}
    \caption{Speedup due to \emph{equalization}, \emph{rasterization}, and \emph{partial enumeration}. All kernels without speedup (gray bars) are not included in the geometric mean. Only few kernels run fast without any optimization (gray labels).\label{fig-optspeedups}}
\end{figure}

\begin{figure*}
  \centering
    \begin{subfigure}{\columnwidth}
    \centering
    \includegraphics[width=\columnwidth]{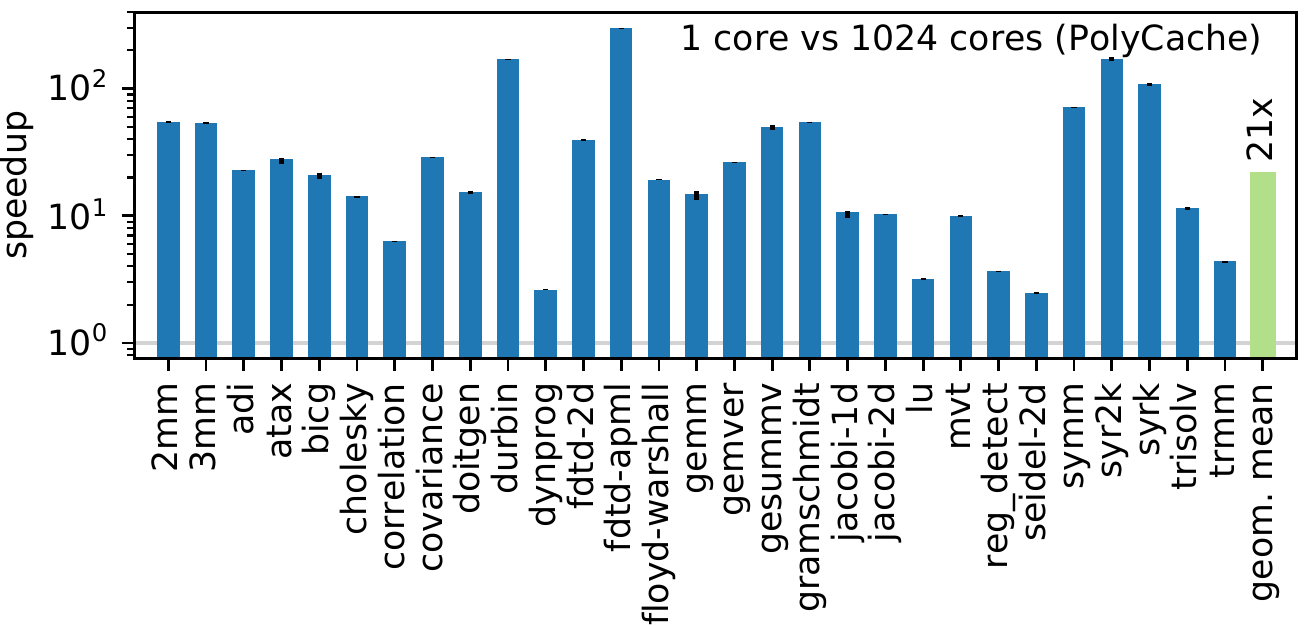}
    \caption{PolyCache\label{fig-benchpolycache}}
    \end{subfigure}\hfill
  \centering
    \begin{subfigure}{\columnwidth}
    \centering
    \includegraphics[width=\columnwidth]{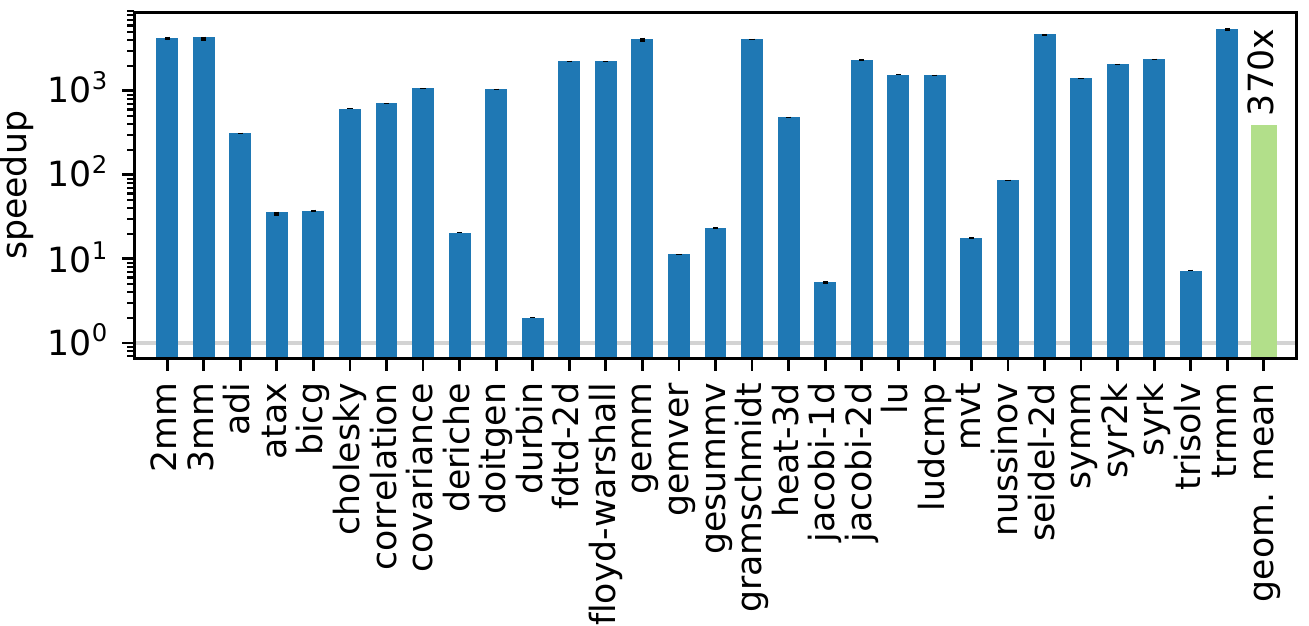}
    \caption{Dinero~IV\label{fig-benchdinero}}
    \end{subfigure}\hfill
    \caption{Speedup of HayStack compared to PolyCache and Dinero for the PolyBench 3.2 and 4.2.1 kernels, respectively.} 
\end{figure*}

Other than for a cache simulator, the model execution time is not proportional to the number of memory accesses. \autoref{fig-benchsizes} shows the model execution times for the three largest PolyBench problem sizes. The large (L) and the extra large (XL) problem size perform roughly 100 and 1000 times more memory access than the medium (M) problem size, respectively. Yet, the execution times remain constant for a majority of the kernels. Only the execution times of the expensive kernels increase since the \emph{partial enumeration} requires more splits. The number of counted pieces, shown by the solid, dashed, and dotted lines in \autoref{fig-benchsizes}, correlate with the cost increase for the larger problem sizes. Even for the expensive kernels, the increase of the execution time is not proportional to the number of memory accesses since we enumerate only the non-affine dimensions of the stack distance polynomials.

When counting the cache misses for multiple cache hierarchy levels, we reuse the stack distance polynomials and enumerate the non-affine dimensions only once. The counting of the individual pieces is the only step repeated for every cache size. 
As the Barvinok algorithm~\cite{barvinok} supports parametric counting, we can count the capacity misses parametric in the cache size which avoids any additional overhead when modeling additional cache hierarchy levels. We benchmark the non-parametric version of the code as it runs faster even when modeling three cache hierarchy levels. \autoref{fig-benchcaches} shows minor increases of the total execution time for two and three cache hierarchy levels.


\newcolumntype{R}[2]{%
    >{\adjustbox{angle=#1,lap=\width-(#2)}\bgroup}%
    l%
    <{\egroup}%
}
\newcommand*\rot{\multicolumn{1}{R{60}{0.3em}}}
\begin{table}
  \caption{Number of non-affine polynomials with zero, one, or two affine dimensions.}
  \begin{tabular}{lcccccccccccc} 
  & \rot{3mm} &
  \rot{adi} &
  \rot{cholesky} &
  \rot{correlation} &
  \rot{covariance} &
  \rot{deriche} &
  \rot{gemver} &
  \rot{lu} &
  \rot{ludcmp} &
  \rot{mvt} &
  \rot{nussinov} &  \\
  \Xhline{2\arrayrulewidth}
  0d-affine & & & 3 & & & & & 4 & 3 & & 7 \\
  \Xhline{1\arrayrulewidth}
  1d-affine & & 7 & 11 & 3 & 3 & 6 & 4 & 48 & 52 & 2 & 18 \\
  \Xhline{1\arrayrulewidth}
  2d-affine & 1 & 59 & 85 & 1 & 1 & & & 27 & 20 & & 48 \\
  \Xhline{2\arrayrulewidth}
  \end{tabular}
\label{tab-affinity}
\end{table}

The \emph{partial enumeration}, discussed in \autoref{sec-capacitymisses}, combines enumeration of the non-affine dimensions with symbolic counting of the affine dimensions. \autoref{fig-optspeedups} compares \emph{partial enumeration} to the explicit enumeration of all points. When considering only kernels with non-affine stack distance polynomials, we measure a geometric mean speedup of 12.4x with pieces that contain 4,400 points on average. The more points per piece the bigger the efficiency gain due to our hybrid counting approach.
We still require explicit enumeration for all non-affine polynomials without affine dimension. \autoref{tab-affinity} shows that most non-affine polynomials have at least one affine dimension. For these polynomials, \emph{partial enumeration} reduces the asymptotic complexity of the capacity miss counting.

As discussed by \autoref{sec-tuning}, 
the floor elimination techniques simplify non-affine stack distance polynomials with less splits than \emph{partial enumeration} but are less generic and do not apply to all polynomials.
\autoref{fig-optspeedups} shows the speedups for \emph{equalization} compared to a baseline without \emph{equalization} and \emph{rasterization}. We disable both techniques since otherwise \emph{rasterization} optimizes the polynomials normally handled by \emph{equalization}. We observe a geometric mean speedup of 1.9x for the kernels that benefit. \autoref{fig-optspeedups} also compares the speedups for \emph{rasterization} to a baseline without \emph{rasterization}. We measure a geometric mean speedup of 1.9x for cholesky, lu, ludcmp, nussinov, and seidel-2d. Overall the floor elimination techniques reduce the number of counted pieces by more than 80\% which results in bigger pieces with better counting performance.

A majority of the kernels perform well independent of problem size and number of cache hierarchy levels. Yet, the model execution times for kernels with non-affine polynomials are higher and problem size dependent. We mitigate this with efficient enumeration and floor elimination techniques.

\subsection{Comparison to PolyCache and Dinero}

The polyhedral cache model PolyCache~\cite{polycache} and the cache simulator Dinero~IV~\cite{dinero} are alternative cache modeling tools. We compare their performance to HayStack.


PolyCache models set associative caches with an LRU replacement policy. We compare to the published results that show the performance for the default problem size of PolyBench 3.2 and adapt the configuration of our model to match the cache sizes of the published experiments (32KiB of L1 cache and 256KiB of L2 cache). 
The only difference is that we model fully associative caches instead of 4-way associative caches.
\autoref{fig-benchpolycache} shows an average speedup of 21x (geometric mean) of HayStack compared to PolyCache even though PolyCache computes the cache misses for all 1024 cache sets in parallel.


Dinero~IV is a trace driven cache simulator which means the expected simulation cost are proportional to the number of memory accesses (\autoref{fig-intro}). \autoref{fig-benchdinero} shows the speedup of HayStack compared to the Dinero~IV simulation times that include the trace generation with QEMU~\cite{qemu}. Dinero~IV simulates the associativity of our test system while we model fully associative caches. As simulation and model run single core, the execution times are comparable. We measure an average speedup of 370x (geometric mean) for the large problem size that would be even bigger for the extra large problem size. Simulating full associativity further increases the average simulation time by factor 2.2x (geometric mean).

PolyCache models cache behavior in-depth, which allows developers to analyze the effects of set associativity and different write policies, but its high accuracy can make it costly to compute. Dinero~IV works for small problem sizes but the cost increase for realistic problem sizes is dramatic.

\subsection{Performance for Tiled Codes}

A tiled code decomposes the iteration domain into tiles and executes tile-by-tile to improve the spacial locality. Tiling can double the loop nest depth which allows us to evaluate our approach for more complex codes. At the same time, estimating the benefits of tiling or even selecting optimal tile sizes is an important application for a cache model.

\begin{figure}
	\centering
	\includegraphics[width=\columnwidth]{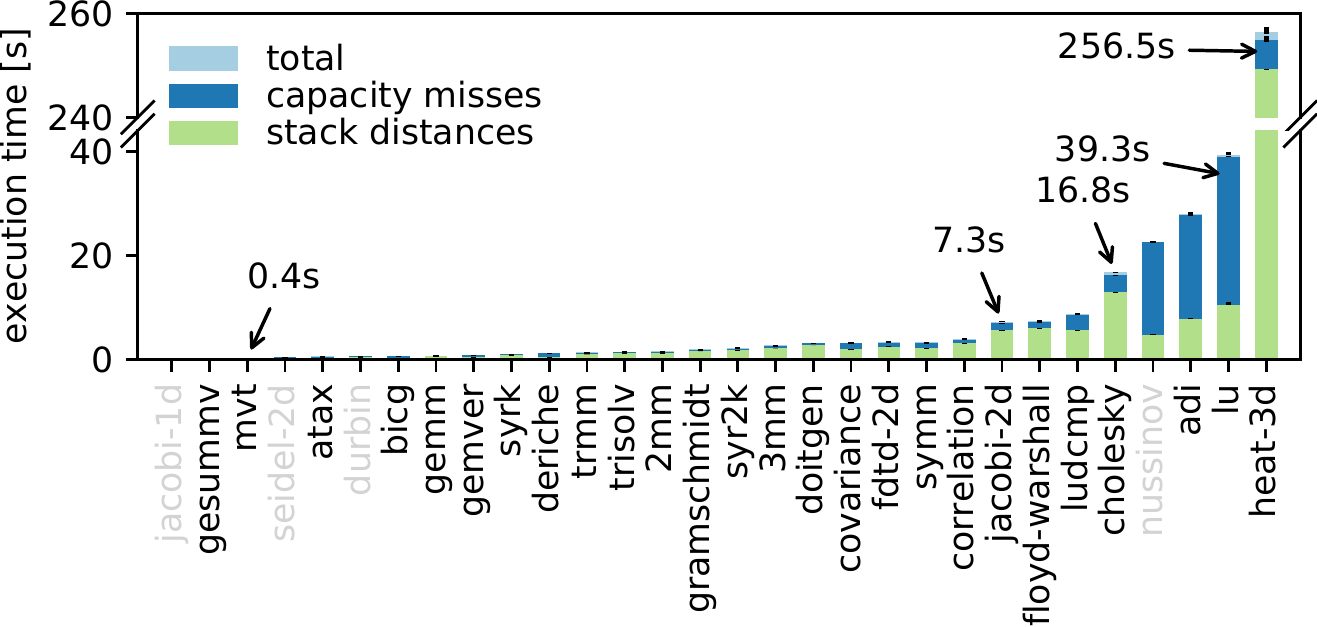}
  \caption{Execution times for the main components of HayStack for tiled versions of the PolyBench kernels. A few kernels (gray labels) have no rectangular tiling.\label{fig-benchtiled}}
\end{figure}

We employ the PPCG~\cite{ppcg} source-to-source compiler to tile all PolyBench kernels with tile size 16. We limit the sum of all scheduling coefficients to one and disable loop fusion to obtain a rectangular tiling without loop skewing (time-tiling). All kernels except for jacobi-1d, durbin, seidel-2d, and nussinov have a rectangular tiling. \autoref{fig-benchtiled} shows the model execution times for the tiled kernels. Tiling makes the cache miss computation more expensive. Especially the stack distance computation of the head-3d kernel runs long. We attribute the cost increase to the more complex iteration domains and memory access patterns.



Tiling increases the model execution times but for a majority of the kernels the cache miss computation still takes only a few seconds.

\section{Related Work}

Cache behavior analysis is a prerequisite when tuning for the memory hierarchy. We distinguish three main approaches: 1) simulation, 2) profiling, and 3) analytical modeling.

\paragraph{Simulators}
Dinero~\cite{dinero} and CASPER~\cite{casper} are examples of trace-based cache simulators that compute the cache misses for the full memory hierarchy. Sniper~\cite{sniper} and gem5~\cite{gem5} have a broader scope and simulate the full system including the caches. All simulators execute the program to count the cache misses which means the simulation costs are proportional to the number of executed memory accesses.

\paragraph{Profiling}
Multiple works discuss the analysis of memory access traces to extract locality metrics. Mattson et al.~\cite{firststackdistance} compute the stack distance using a linked list and derive the cache hit rate for different cache sizes. Tree based implementations~\cite{bennett1975lru,olken1981efficient,sugumar1993efficient} reduce the cost of the stack distance computation. Kim et al.~\cite{kim1991implementing} apply hashing and approximation to increase the efficiency.
Ding et al.~\cite{ding2003predicting} discuss tree based approximate algorithms that reduce the time and space complexity of the stack distance computation and predict the stack distance histogram for arbitrary problem sizes given training inputs for few different problem sizes.
Eklov et al.~\cite{eklov2010statstack} sample the reuse distance for a few memory accesses and employ statistics to estimate stack distances and cache miss ratio.
Xiang et al.~\cite{xiang2013hotl} discuss five different locality metrics and show how to derive miss rate and reuse distance given the a single measure called average footprint which they compute with an efficient linear time algorithm~\cite{xiang2011linear}.
A disadvantage of the profiling approaches is the acquisition and the handling of the large program traces. 
Chen et al.~\cite{chen2018locality} sample the reuse time during compilation which allows them to estimate the cache miss ratio of complex loop nests.

\balance

\paragraph{Analytical models}
Agarwal et al.~\cite{agarwal1989analytical} develop an analytical model that uses parameters extracted from the program trace. Harper et al.~\cite{harper1999analytical} model set associative caches for regular loop nests. Cost models~\cite{kennedy1992optimizing, carr1994compiler, bondhugula2008practical} allow compilers to decide if data-locality transformations are beneficial. All of these models only approximate the number of cache misses.

Ferdinand et al.~\cite{agebasedai} use abstract interpretation to model set associative LRU caches. Model-checking~\cite{modelcheck1, modelcheck2} increases the accuracy of this analysis that distinguishes always hit, always miss, and not classified. Touzeau et al.~\cite{exactai} show how to attain high accuracy without costly model-checking. The abstract interpretation approaches are complementary to our cache model since they support dynamic control flow but approximate the cache misses of loop nests by classifying all instances of a memory access at once. 

Ghosh et al.~\cite{ghosh1998precise} derive cache miss equations to count the cache misses for perfect loop nests with data dependencies represented by reuse vectors~\cite{reusevector}. Assuming an LRU replacement policy, a cache miss occurs if the number of solutions to a cache miss equality exceeds the cache associativity. Counting the solutions for every point of the iteration domain is expensive. Vera and Xue et al.~\cite{vera2002let, xue2004efficient} thus sample the iteration domain to speedup the cache miss computation which allows them to perform approximate whole-program analysis.
Cascaval et al.~\cite{cabetacaval2003estimating} compute the stack distance histogram symbolically for perfect loop nests with uniform data dependencies. They model fully associative caches with an LRU replacement policy and use statistics to model set associative caches.
Chatterjee et al.~\cite{chatterjee2001exact} use Presburger formulas to express the set of compulsory and capacity misses of imperfect loop nests for associative caches. At the time, their approach was limited to small problem sizes and low associativity since the computation of analytical results for realistic hardware and even small benchmarks kernels was prohibitively complex.
%
While Beyles et al.~\cite{epic} did not address the cache miss problem, they use analytically computed stack distance to generate cache hints at runtime. Their stack distance computation, extended by our cache miss counting technique for non-affine polynomials, is the foundation of our cache model.
PolyCache~\cite{polycache} presented the first analytical approach fast enough
to compute the cache behavior of static control programs for interesting
benchmark kernels and realistic hardware parameters. Its analytical model
relates for every cache set successive accesses of distinct cache lines and
repeatedly removes the shortest relations to model set associativity with LRU
replacement policy. While PolyCache also uses symbolic counting techniques to
avoid a complete enumeration of the computation, its complexity increases with
high associativity. Our work provides a fast analytical model for fully
associative caches and shows that fully associative models introduce only small errors compared to measurements on actual hardware.

\section{Conclusion}

As memory behavior depends on the cache state, understanding the cost of
memory accesses is much more difficult than understanding the cost
of arithmetic instructions. With HayStack, we close this gap by providing
developers with accurate information about the interaction of memory accesses
with the large and deep cache hierarchy of modern processors. HayStack allows the programmer
to predict memory access costs accurately and to develop programs well
optimized for the memory hierarchy. When striving for ultimate performance,
both a good baseline and an accurate surrogate model accelerates empirical
tuning. As a result, cache-aware program optimization becomes accessible.

Responsiveness is key for the adoption of any cache model. We demonstrate
excellent often problem size independent response times that for the
first time make analytical cache modeling practical. 
%
%
In addition, the cache
size independent costs allow our model to easily scale to future hardware.
We show the practicality of our deliberate decision against high fidelity and in favor of a generic fully associative cache model. 
%
The proposed model is robust to memory layout choices and hardware
implementation details and yet reaches very high accuracy on real hardware across
a wide range of computations.

\begin{acks}     

This project has received funding from the European Research Council (ERC)
under the European Union's Horizon 2020 programme (grant agreement DAPP, No.
678880), the Swiss National Science Foundation under the Ambizione programme (grant PZ00P2168016), and ARM Holdings plc and Xilinx Inc in the context of Polly Labs. We also would like to thank the Swiss National Supercomputing Center for providing the computing resources.

\end{acks}


\bibliography{references}

\end{document}